\documentclass[12pt]{iopart}
\usepackage{graphicx}

\usepackage{bm}%
\usepackage[T1]{fontenc}
\expandafter\ifx\csname package@font\endcsname\relax\else
 \expandafter\expandafter
 \expandafter\usepackage
 \expandafter\expandafter
 \expandafter{\csname package@font\endcsname}%
\fi
\ifx\pdfoutput\undefined
\usepackage{graphicx}
\else
\fi
\usepackage{hyperref}

\hypersetup{
    bookmarks=false,         
    unicode=false,          
    pdftoolbar=false,        
    pdfmenubar=false,        
    pdffitwindow=false,     
    pdfstartview={FitH},    
    pdfnewwindow=true,      
    colorlinks=true,       
    linkcolor=black,          
    citecolor=blue,        
    filecolor=blue,      
    urlcolor=blue           
}

\usepackage{natbib}
\bibpunct{(}{)}{,}{s}{,}{,}

\begin{document}

\title{Input-output description of microwave radiation in the dynamical Coulomb blockade}

\author{J~Lepp\"akangas$^1$, G~Johansson$^1$, M~Marthaler$^2$ and M~Fogelstr\"om$^1$  }

\address{$^1$ Microtechnology and Nanoscience, MC2, Chalmers
University of Technology, SE-412 96 G\"oteborg, Sweden}
\address{$^2$ Institut f\"ur Theoretische Festk\"orperphysik
and DFG-Center for Functional Nanostructures (CFN), Karlsruhe Institute of Technology, D-76128 Karlsruhe, Germany}
\ead{juha.leppakangas@chalmers.se}

\begin{abstract}
We study microwave radiation emitted by a small voltage-biased Josephson junction connected
to a superconducting transmission line. An input-output formalism for the radiation field is established,
using a perturbation expansion in the junction's critical current.
Using output field operators solved up to the second order, we estimate the spectral density and the second-order coherence of the emitted field.
For typical transmission line impedances and at frequencies below the main emission peak at the Josephson frequency, radiation occurs predominantly due to two-photon emission.
This emission is characterized by a high degree of photon bunching if detected symmetrically around half of the Josephson frequency.
Strong phase fluctuations in the transmission line make related nonclassical phase-dependent amplitude correlations short lived, and there is no steady-state two-mode squeezing.
However, the radiation is shown to violate the classical Cauchy-Schwarz inequality of intensity cross-correlations,
demonstrating the nonclassicality of the photon pair production in this region.
\end{abstract}



\maketitle

\section{Introduction}

In a resistive environment, charge tunneling across a voltage-biased Josephson junction (JJ) triggers simultaneous microwave 
emission~\cite{Devoret1990,Ingold1992}, that carries away all or some of the gained electrostatic energy.
For voltages below the superconducting gap, $eV<\Delta$, the radiation is purely due to Cooper-pair tunneling since charge tunneling
via excitation of quasiparticles is energetically forbidden.
This mechanism leads to spectroscopically sharp features, which can be used for probing of environmental energy levels~\cite{Holst1994,Leppakangas2006,SaclayNature2013}, 
as photon absorption by the environmental modes influences the simultaneously measurable dc-current.
A novel idea is to use the voltage-biased small JJ as a source of nonclassical microwave radiation, i.e.~to convert the applied dc-voltage
into correlated microwave photons~\cite{Hofheinz2011,Leppakangas2013}.
This has  recently stimulated theoretical studies of the emitted microwave field~\cite{Leppakangas2013,Nazarov,Ulm,Nottingham}.

In this article, we investigate microwave radiation produced in a dc-voltage-biased superconducting transmission line that is terminated by a small JJ.
We establish an input-output formalism for the field operators in the transmission line.
In this formalism, the electric current across the Josephson junction acts as a nonlinear and time-dependent boundary condition for the
microwave field~\cite{Johansson2010,WallquistPRB}. We solve this perturbatively as a power series in the junction's critical current~\cite{Leppakangas2013}, 
and give explicit expressions for the output field operators up to the second order in the critical current.
Assuming thermal equilibrium of the input field we recover the limit of incoherent Cooper-pair tunneling~\cite{Devoret1990,Ingold1992},
where microwave emission is due to an incoherent sequence of Cooper-pair tunneling events.
The photon flux has been studied recently experimentally in this regime,~\cite{Hofheinz2011}
and it was shown that the radiation at voltages below the Josephson frequency, for typical tranmission-line impedances, is due to simultaneous two-photon emission.
Using the formalism established in this article, we also study the nonclassicality of the photon pair production occurring in this region.

Using field operators up to second order in the junction's critical current,
we derive analytical expressions for the first and and second-order (photon) coherences
for typical transmission lines.
We reproduce results for the photon-flux density and simultaneous electric current,
previously derived using the $P(E)$-theory~\cite{Ingold1992,Hofheinz2011}.
Further, the emission characteristics below the Josephson frequency is shown to be highly bunched,
if detected symmetrically around half the Josephson frequency, $eV/h$.
We then further study the nonclassicality of the photon pair production occuring below the Josephson frequency~\cite{Leppakangas2013}.
Strong phase fluctuations in the transmission line
make phase-dependent nonclassical amplitude correlations short lived,
and lead to a vanishing two-mode squeezing, in the steady-state.
We thus proceed to prove the nonclassicality of the radiation in a different way,
considering the classical Cauchy-Schwarz inequality of intensity cross-correlations,
a nonclassicality test that is not affected by dephasing.
Using the developed methods, we derive an equivalent inequality but expressed
in terms of $P(E)$-functions. This is used to show that
the emitted photons below the Josephson frequency violate the inequality,
demonstrating the nonclassicality of the photon pair production in this region.

The article has the following structure. In Section 2 we introduce the model we use to describe the radiation in the transmission line,  and the nonlinear and time-dependent boundary condition created by the Josephson junction.
In section 3, we derive the solution by establishing a perturbation expansion in the junction's critical current.
In section 4, we derive results for the microwave spectral density
in the used leading-order approximation, whose validity is also addressed in this section.
Higher-order coherences and the nonclassicality of the output radiation are investigated in Section 5.
The technical details of the calculations are given in Appendices A-D.

\section{The system and the model}
\begin{figure}[bt]
\includegraphics[width=0.9\linewidth]{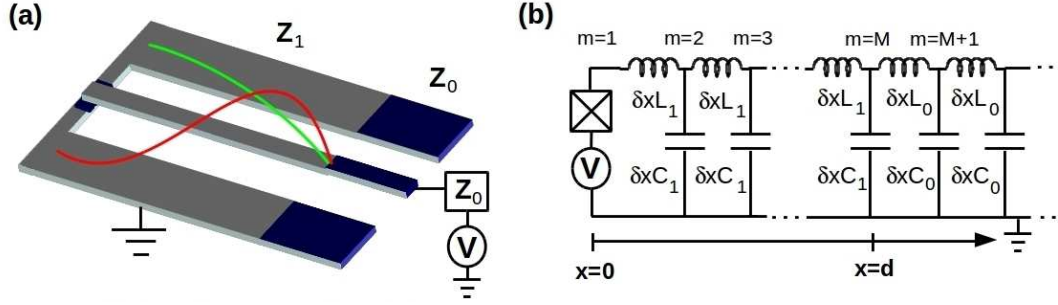}
\caption{(a) We consider a dc-voltage-biased transmission line terminated by a small Josephson junction.
We allow for a step-like change in the characteristic impedance from $Z_0$ to $Z_1$, supporting the depicted modes of a $\lambda/4$ resonator.
(b) The equivalent circuit model consists of an infinite series of capacitors $\delta x C_i$ and inductors $\delta x L_i$,
where $\delta x$ is a small discretization width, that is let to approach zero ($\delta x M=d$),
in parallel with a Josephson junction (crossed box). We have $Z_i=\sqrt{L_i/C_i}$.
The upper conductor consists of series of islands, each assigned with a counting parameter $m$.
}
\label{figure1}
\end{figure}

Our system consists of a dc-voltage-biased transmission line terminated by a small Josephson junction, figure~\ref{figure1}(a).
We consider explicitly two types of environments:
(i) a semi-infinite transmission line (i.e.~$Z_0=Z_1$) and
(ii) a semi-infinite transmission line with a  $\lambda/4$ cavity (i.e.~$Z_1 > Z_0$). 
Case (i) allows for analytical solutions, while case (ii) enhances the output radiation at the cavity resonances, which is important in experiments.


\subsection{Heisenberg equations of motion and quantization of the EM field}
Following~\cite{WallquistPRB}, we model the system using a discretized circuit model, depicted in figure~\ref{figure1}(b).
The total Lagrangian of the system can be decomposed as
\begin{equation}
{\cal L}={\cal L}_{\rm c}+{\cal L}_{\rm f}+{\cal L_{\rm J}}.
\end{equation}
The cavity ($0<x<d$) and the free space ($x>d$) Lagrangians are respectively~\cite{Johansson2010}
\begin{eqnarray}
{\cal L}_{\rm c}&=&\sum_{m \geq 1}^M\frac{\delta x C_1\dot\Phi_m^2}{2}-\sum_{m\geq 2}^N\frac{(\Phi_m-\Phi_{m-1})^2}{2L_1\delta x},\\
{\cal L}_{\rm f}&=&\sum_{m>M}\frac{\Delta x C_0\dot\Phi_m^2}{2}-\sum_{m\geq M}\frac{(\Phi_m-\Phi_{m+1})^2}{2L_0\Delta x}.
\end{eqnarray}
Here $\Phi_m(t)$ is the magnetic flux of node $m$, see figure~\ref{figure1}(b).
The Josephson junction is described by the term
\begin{equation}
{\cal L_{\rm J}}=\frac{ C_{\rm J}\dot\Phi_1^2}{2}+E_{\rm J}\cos\left( 2\pi \frac{\Phi_1-\Phi_V}{\Phi_0}\right).
\end{equation}
Here $C_{\rm J}$ is the junction's capacitance, $E_{\rm J}$ is the Josephson coupling energy, $\Phi_0=h/2e$ is the
flux quantum, and the dc-voltage bias results in the term $\Phi_V=Vt$.

We consider the Heisenberg equations of motion in the continuum limit $\delta x\rightarrow 0$.
Inside each region, free space ($i=0$) or cavity ($i=1$), one obtains a Klein-Gordon equation,
\begin{equation}
\ddot\Phi(x,t)=\frac{1}{L_iC_i}\frac{\partial^2\Phi(x,t)}{\partial^2 x}.
\end{equation}
Here $\Phi(x,t)$ is the position dependent magnetic flux. We write down a solution for the cavity region as ($0<x<d$)
\begin{equation}
\Phi(x,t)=\sqrt{\frac{\hbar Z_{1}}{4\pi}}\int_0^\infty\frac{d\omega}{\sqrt{\omega}}\times\left[ a_{\rm  in}^{\rm c}(\omega)e^{-i(k_\omega^{\rm c} x+\omega t)}+a_{\rm out}^{\rm c}(\omega)e^{-i(-k_\omega^{\rm c} x+\omega t)}+{\rm H.c.} \right]. 
\end{equation}
Here $Z_1=\sqrt{L_1/C_1}$ is the corresponding characteristic impedance and $k_{\omega}^{\rm c}=\omega\sqrt{C_1L_1}$ the wave number. Similarly, we write the free-space solution as ($x>d$)
\begin{equation}
\Phi(x,t)=\sqrt{\frac{\hbar Z_{0}}{4\pi}}\int_0^\infty\frac{d\omega}{\sqrt{\omega}}\times \left[ a_{\rm  in}^{\rm f}(\omega)e^{-i(k_\omega^{\rm f} x+\omega t)}+a_{\rm out}^{\rm f}(\omega)e^{-i(-k_\omega^{\rm f} x+\omega t)}+{\rm H.c.} \right].  
\end{equation}

The in-field
creation operators of photons, $a^{\dagger}(\omega)$, and the annihilation operators, $a(\omega)$,
satisfy the commutation relation~\cite{Peskin}
\begin{equation}\label{CommutationRelation}
\left[ a_{\rm in}(\omega),a_{\rm in}^{\dagger}(\omega') \right]=\delta(\omega-\omega').
\end{equation}
As a consistency check of our theory, we will show that (\ref{CommutationRelation}) is satisfied also for the out field operators.

\subsection{Boundary conditions for the EM field}
The boundary conditions appear at the Josephson junction ($x=0$) and at the possible
discrete change of the transmission-line parameters ($x=d$).
Generally, we will have three boundary conditions to solve, and three unknown fields [$a_{\rm in}^{\rm c}(\omega)$, $a_{\rm out}^{\rm c}(\omega)$, and $a_{\rm out}^{\rm f}(\omega)$].
The requirements of a continuous voltage distribution and current conservation across $x=d$ imply the linear conditions,
\begin{eqnarray}
\Phi(d^-,t)&=&\Phi(d^+,t), \label{eq:BoundaryLine1} \\
\frac{\partial\Phi(d^-,t)}{L_0\partial x}&=&\frac{\partial\Phi(d^+,t)}{L_1\partial x} .\label{eq:BoundaryLine2}
\end{eqnarray}
These can be solved by Fourier transformation.

The main challenge is to solve the boundary condition at the junction, where  current
conservation gives a nonlinear and time-dependent condition
\begin{equation}\label{RCJ}
C_{\rm J}\ddot\Phi(0,t)+\frac{1}{L_0}\frac{\partial\Phi(x,t)}{\partial x}\vert_{x=0}=-I_{\rm c}\sin\left[\frac{2\pi}{\Phi_0}\Phi(0,t)-\omega_{\rm J} t\right]  .
\end{equation}
Here, $I_{\rm c}=(2\pi/\Phi_0)E_{\rm J}$ is the junction's critical current and $\omega_{\rm J}=2eV/\hbar$ is  the Josephson frequency.
Classically, for $Z_0=Z_1$ and $T=0$ (no input), this is equivalent to RCJ-model of a Josephson junction~\cite{Likharev}.


\section{Perturbative input-output approach}
We  want to derive a solution for the outgoing free-space field, $a^{\rm f}_{\rm out}(\omega)$, as a function of the incoming field in the same region, $a^{\rm f}_{\rm in}(\omega)$.
The challenge is that the boundary condition at the junction is highly nonlinear,
containing all moments of the field operators $a^{\rm c}_{\rm in/out}(\omega)$ and $\left[a^{\rm c}_{\rm in/out} (\omega) \right]^{\dagger}$.
A linearization of this condition, performed in Appendix A, captures some of the essential physics,
but is limited to the second order in the photonic processes and does not, for example, describe correctly the effect of low-frequency phase fluctuations.
In the case of a weakly damped resonator ($Z_1\gg Z_0$), a single cavity mode can also be picked out and be described as a damped oscillator~\cite{Ulm,Nottingham}, but  with a limited description of the important low-frequency modes.
Here, we use a different approach and derive a solution for the continuous-mode output field operators as a power series in the junction's critical current $I_{\rm c}$.

\subsection{Unperturbed solution}
The starting point is the solution for $I_{\rm c}=0$, i.e.~when Cooper-pair tunneling is neglected.
By a Fourier transformation (Appendix B) we obtain the linear dependence
\begin{eqnarray}\label{ZerothSolutionMain}
a_{\rm out}^{\rm f}(\omega)&=&R (\omega) a_{\rm in}^{\rm f}(\omega)\\
R (\omega)&=&\frac{1+re^{-2ik^{\rm c}_\omega d-i\theta(\omega)}}{e^{-i\theta(\omega)}+re^{2ik^{\rm c}_\omega d}}  . \nonumber
\end{eqnarray}
Here  $e^{i\theta(\omega)}={\cal C}^* (\omega)/{\cal C}(\omega)$, ${\cal C}(\omega)=1+iZ_1C_{\rm J}\omega $, and $r = (Z_1-Z_0)/(Z_0+Z_1)$.
Similarly, we can solve for the cavity out-field $a_{\rm out}^{\rm c}(\omega)$ as a
function of the free-space input $a_{\rm in}^{\rm f}(\omega)$, 
\begin{eqnarray}\label{ResonanceStructure}
a_{\rm out}^{\rm c}(\omega)&=&A(\omega)a_{\rm in}^{\rm f}(\omega) \\
A(\omega)&=&\frac{2\sqrt{\frac{Z_1}{Z_0}} e^{-2ik^{\rm c}_{\omega}d}}{\left(1+\frac{Z_1}{Z_0} \right)e^{-2ik_\omega^{\rm c} d-i\theta(\omega)}+\frac{Z_1}{Z_0}-1}, \nonumber
\end{eqnarray}
where $A(\omega)$ gives the response of the cavity to external drive and
possesses information of its resonance frequencies.
With the help of this,
the operator for the phase difference at the junction, $\phi_0(t)\equiv 2\pi\Phi(0,t)/\Phi_0$, can be written
\begin{eqnarray}\label{PhaseDifference0}
\phi_0(t)=\frac{\sqrt{4\pi\hbar Z_1}}{\Phi_0}\int_{0}^{\infty}  \frac{d\omega}{\sqrt{\omega}} \bar A(\omega) a_{\rm in}^{\rm f}(\omega)e^{-i\omega t}+{\rm H.c.}
\end{eqnarray}
Here, $\bar A(\omega)=A(\omega)/{\cal C}^*(\omega)$ and we also note the useful relation $R(\omega)=\bar A(\omega)/\bar A^*(\omega)$.
The corresponding phase fluctuations are equivalent to that of  the "tunneling" impedance $Z_{\rm t}(\omega)$~\cite{Ingold1992}, defined as
\begin{eqnarray}
{\rm Re}[Z_{\rm t}(\omega)]&\equiv& Z_1\vert \bar A(\omega)\vert^2, \label{EffectiveImpedance} \\
\left\langle\phi_0(t)\phi_0(t')\right\rangle&=&2\int_{-\infty}^{\infty}\frac{d\omega}{\omega}\frac{{\rm Re}[Z_{\rm t}(\omega)]}{R_Q}\frac{e^{-i\omega(t-t')}}{1-e^{-\beta\hbar\omega}},\nonumber
\end{eqnarray}
where $R_{\rm Q}=h/4e^2$ is the (superconducting) resistance quantum.
In the ensemble average we assume thermal equilibrium for the incoming free-space modes,
used throughout this article.
For the open line ($Z_0=Z_1$) case, the impedance~(\ref{EffectiveImpedance}) describes a  capacitor $C_{\rm J}$ and a resistor $Z_0$ in parallel, whereas
for the cavity case ($Z_1\gg Z_0$) it describes a capacitively shunted $\lambda/4$-resonator, with resonances approximately
at $\omega_k=(2k+1)\omega_0$, where $\omega_0=\pi/2 d\sqrt{C_1L_1}$ and $k\in[0,1,2,\ldots ]$.

\subsection{First and second order solutions}
To seek for a solution, that includes Cooper-pair tunneling ($I_{\rm c}\neq 0$), we multiply the right-hand side of boundary condition~(\ref{RCJ}) by a formal dimensionless parameter $\xi$, and correspondingly write the
solution for the annihilation operators of the outgoing field in the open space as $a_{\rm out}^{\rm f}(\omega)=\sum_{n=0}^{\infty}\xi^n a_n(\omega)$. The zeroth-order solution, $a_0(\omega)$, corresponds to $I_{\rm c}=0$ and was obtained in (\ref{ZerothSolutionMain}).
We make a similar expansion for the fields inside the cavity and for the phase difference across the Josephson junction.
The input field in the free space, $a_{\rm in}^{\rm f}(\omega)$,
is  independent of $\xi$, as the output in this region does not reflect back.
The task is to find the other fields as a function of the known input, $a_{\rm in}^{\rm f}(\omega)$, for small $I_{\rm c}\neq 0$.

We solve the boundary condition at the junction order by order in $\xi$. 
By a straightforward calculation we find the leading-order solution  (Appendix B)
\begin{eqnarray}\label{SolutionOut1}
a^f_{1}(\omega)=i  I_{\rm c} \sqrt{\frac{Z_1}{\hbar\omega\pi}} \bar A(\omega) \int_{-\infty}^{\infty}dt e^{i\omega t} \sin\left[  \phi_0(t)-\omega_{\rm J} t \right] .
\end{eqnarray}
We observe that the operator is a (sinusoidal) function of the zeroth-order phase-difference operator (\ref{PhaseDifference0}).
In the second order for $\xi$ (and $I_{\rm c}$) we obtain
\begin{eqnarray}\label{SolutionOut2}
a^f_{2}(\omega)=i  I_{\rm c} \sqrt{\frac{Z_1}{\hbar\omega\pi}} \bar A(\omega) \int_{-\infty}^{\infty}dt e^{i\omega t}  \left[ \sin\left[  \phi_0(t)-\omega_{\rm J} t \right],z(t) \right] .
\end{eqnarray}
Here, the  operator $z(t)$ ($\propto I_{\rm c}$) is a solution to the equation $\phi_1(t)=[\phi_0(t),z(t)]$,
where $\phi_1$ is the phase-difference operator in the first order, obtained via the leading-order solution (\ref{SolutionOut1}), see Appendix B.  Important here is that the operators $\phi_i(t)$ do not commute with each other.

Thus, we have obtained a solution for (\ref{RCJ}) to second order in $\xi$ as a function of the operator $z(t)$, which still needs to be solved.
In the case of semi-infinite transmission line ($Z_0=Z_1$), we find a simple explicit form of $z(t)$, 
\begin{eqnarray}
z(t)=-i\frac{E_{\rm J}}{2\hbar}\int_{-\infty}^{\infty}dt'\left[ 1+\frac{{\rm Sgn}(t-t')}{e^{\omega_{\rm c}\vert t-t'\vert}-1} \right] \cos[ \phi_0(t')-\omega_{\rm J} t' ].\nonumber
\end{eqnarray}
Here, the junction RC-time gives the high frequency cut-off $\omega_{\rm c}=1/Z_0C_{\rm J}$. The solution has an apparent divergence
at $t=t'$, which cancels for symmetry reasons in all measurable quantities discussed in this article.
We observe that the operator $z(t)$ is also a trigonometric function of the zeroth-order phase-difference operator. Also
generally ($Z_0\neq Z_1$), $z(t)$ has the form $\int_{-\infty}^{\infty}dt' {\cal S}(t-t') \cos[ \phi_0(t')-\omega_{\rm J} t' ]$, where ${\cal S}(t-t')$ is a scalar function. 
The trigonometric functions can be decomposed into exponential operators $e^{\pm i[\phi_0(t)-\omega_{\rm J}t]}$, that correspond to charge transfers of $2e$
across the JJ in the two possible directions.~\cite{Ingold1992} Thus, we see that the solutions 
(\ref{SolutionOut1}-\ref{SolutionOut2}) include all possible tunneling processes up to second order.

We can now study the consistency of our solution, by checking if the output radiation field satisfies the commutation relation (\ref{CommutationRelation}). This property is vital as it, for example, secures causality in the theory~\cite{Peskin}.
We obtain up to second order,
\begin{eqnarray}
&&\left[ a^{\rm f}_{\rm out}(\omega),\left[ a^{\rm f}_{\rm out}(\omega')\right]^{\dagger} \right]=\vert R(\omega)\vert^2\delta(\omega-\omega') \nonumber \\
&+& \xi\left( [a_1(\omega),a_0^{\dagger}(\omega')] + [a_0(\omega),a_1^{\dagger}(\omega')] \right)\nonumber \\
&+&  \xi^2\left( [a_2(\omega),a_0^{\dagger}(\omega')] + [a_0(\omega),a_2^{\dagger}(\omega')] + [a_1(\omega),a_1^{\dagger}(\omega')] \right), \label{CommutationRelation2}
\end{eqnarray}
using the input field commutation relation $\left[ a^{\rm f}_{\rm in}(\omega),  \left[a^{\rm f}_{\rm in} (\omega') \right]^{\dagger} \right]=\delta(\omega-\omega')$.
Since $\vert R(\omega)\vert=1$, the first term on the right-hand side produces the desired $\delta$-function. A straightforward calculation (Appendix B) shows that the
rest of the terms, order by order in $\xi$, sum to zero. This confirms that the commutation relation is indeed valid, up to the order our solution allows us to check this.


\begin{figure}[bt]
\includegraphics[width=\linewidth]{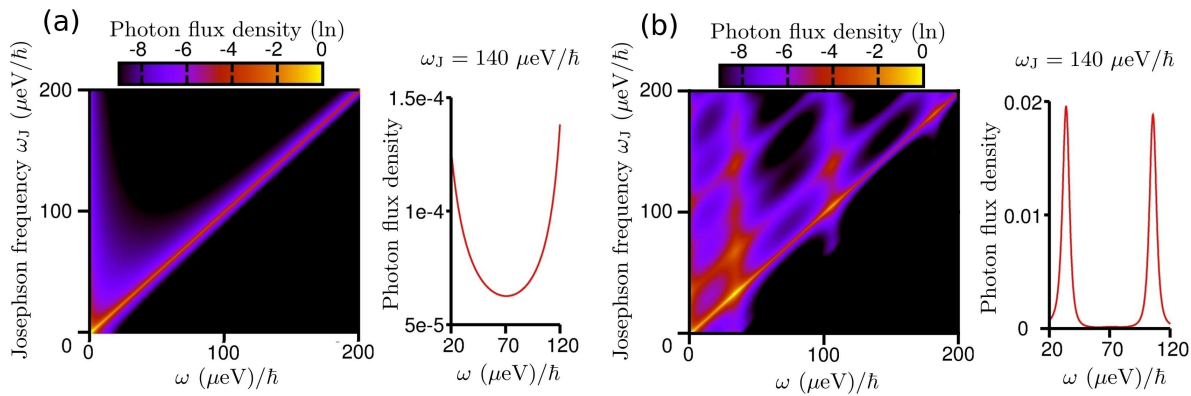}
\caption{Photon-flux density $f_{\rm t}(\omega)$ as a function Josephson frequency $\omega_{\rm J}=2eV/\hbar$ [left panels of (a) and (b)] and for
a single value $\omega_{\rm J}=140$~$\mu$eV/$\hbar$ (right panels). The Josephson radiation is seen as a diagonal resonance $\omega=\omega_{\rm J}$ in the photon flux density (left panels).
The flux is asymmetric around the diagonal, as
for $\omega>\omega_{\rm J}$ the emission is suppressed by the temperature, but for $\omega<\omega_{\rm J}$ multi-photon production results in extra emission. When the pair production of photons dominates, the photon flux becomes symmetric around
half of the Jospehson frequency $\omega_{\rm J}$, as seen for the single picked value $\omega_{\rm J}=140$~$\mu$eV/$\hbar$ (right panels).
In (b) this is approximately the sum of the two resonance frequencies in the cavity,
and emission to these modes is enhanced.
 We use $E_{\rm J}=15$~$\mu$eV, $C=10$~$f$F, $T=100$~mK,
 $Z_0=100$~$\Omega$, (a) $Z_1=100$~$\Omega$ or (b) $Z_1=500$~$\Omega$ with
$f_0=1/4 d\sqrt{C_1L_1}=10$~GHz.
}
\label{figure2}
\end{figure}

\section{Emission characteristics I: Photon-flux density}
Thus, having derived explicit expressions for the outgoing field-operators to second order in the junction's critical current,  we go on to study  properties of the output radiation.
We first investigate general relations for the amplitude correlations, and after this consider their explicit forms.
The truncation of the power series to the leading order can be made for small transparency JJs, i.e.~for small $I_{\rm c}$. The exact definition of
"small" is then addressed in Section \ref{sec:convergence}. In later parts of the article, Sections \ref{SecondOrderCoherence} and \ref{Nonclassicality}, we discuss results of similar calculations but done for higher-order correlations.

\subsection{General properties for the amplitude correlations}
By a direct calculation, we obtain for the amplitude correlations related to the photon-flux and power-spectral densities,
\begin{eqnarray}\label{eq:PhotonFluxGeneral}
\left\langle a^{\dagger}_{\rm out}(\omega)a_{\rm out}(\omega') \right\rangle = 2\pi f(\omega)\delta(\omega-\omega').\label{eq:Amp1}
\end{eqnarray}
We use here the notation $ a_{\rm out}(\omega) \equiv a^{\rm f}_{\rm out}(\omega) $. The function $f(\omega)$
is identified as the the photon-flux density~\cite{Loudon}.
This diagonal form is a result of the finite phase-coherence time, present already in the zeroth-order phase-difference~(\ref{PhaseDifference0}).
The phase difference performs a quantum Brownian motion in time,~\cite{Grabert1988} and it follows that expectation values of type $\left\langle e^{i\phi_0(t)} e^{i\phi_0(t')}\right\rangle$ are zero. This also implies that the amplitude correlations related to possible squeezing are zero,
\begin{eqnarray}
\left\langle a_{\rm out}(\omega)a_{\rm out}(\omega') \right\rangle = 0.\label{eq:Amp2}
\end{eqnarray}
In general, due to the random phase fluctuations there is on average no phase coherence in the output radiation.
Further, only even powers of the critical current are present in the power series of the photon-flux density,
or of any higher-order correlator considered in this article,
\begin{equation}
f(\omega)=\sum_{n=0}^{\infty}I_{\rm c}^{2n}{\cal F}_{n}(\omega),
\end{equation}
where functions ${\cal F}_n(\omega)$ are independent of $I_{\rm c}$.
This follows again from the phase fluctuations, namely because $\left\langle \Pi_m e^{\pm i\phi_0(t_m)}\right\rangle=0$,
for odd integers $m$.

We further divide the leading-order result for the output photon-flux-density,
(\ref{eq:PhotonFluxGeneral}), as
\begin{equation}\label{FluxGeneral}
f(\omega)=f_{\rm t}(\omega)+f_{\rm th}(\omega).
\end{equation}
Here the term $f_{\rm t}(\omega)$ is the photon-flux density created by inelastic Cooper-pair tunneling,
and the part $f_{\rm th}(\omega)$ describes reflection of the incoming thermal photons,
being finite only for $T\neq 0$. 
In the following we consider the explicit forms of $f_{\rm t}(\omega)$ and $f_{\rm th}(\omega)$.

\subsection{Emission from inelastic Cooper-pair tunneling $f_{\rm t}(\omega)$}
Radiation due to  inelastic Cooper-pair tunneling is obtained by inserting the leading-order solution for both operators $a^{(\dagger)}_{\rm out}$ in (\ref{eq:Amp1}). A straightforward calculation gives (Appendix C),
\begin{eqnarray}\label{OutputFlux}
&&f_{\rm t}(\omega)= \int_0^{\infty} d\omega '\frac{1}{2\pi}\langle a^{\dagger}_{1}(\omega)a^{}_{1}(\omega') \rangle\\
&=&\frac{I_{\rm c}^2  {\rm Re}[Z_{\rm t}(\omega)] }{2\omega}   \left[P(\hbar\omega_{\rm J}-\hbar\omega)+P(-\hbar\omega_{\rm J} -\hbar\omega)\right]. \nonumber
\end{eqnarray}
Here, the function $P(E)$ is the probability to exchange energy $E$ with the electromagnetic environment, in this case
with the transmission line, defined as
\begin{eqnarray}\label{PEDefinition}
P(E)=\frac{1}{2\pi\hbar}\int_{-\infty}^{\infty}dte^{J(t)+i\frac{E}{\hbar}t},
\end{eqnarray}
where the phase correlator function, $J(t)=\left\langle [\phi_0(t)-\phi_0(0)]\phi_0(0) \right\rangle$, is a measure of phase fluctuations in the zeroth order.
Equation~(\ref{OutputFlux}) was obtained  first in reference~\cite{Hofheinz2011}
by applying the theory of inelastic Cooper-pair tunneling~\cite{Ingold1992}, i.e. $P(E)$-theory, and keeping track of the simultaneously emitted photons.


We will now analyze the obtained photon-flux density in more detail and compare it with the classical solution, which consists
of continuous radiation at the Josephson frequency $\omega_{\rm J}$, broadened by low-frequency
phase fluctuations.
The classical power spectral density, defined  as $S(\omega)=\hbar\omega f(\omega)$, has the approximative form~\cite{Likharev}
\begin{equation}
S_{\rm cl}(\omega)=\frac{\hbar I_{\rm c}^2{\rm Re}[Z_{\rm t}(\omega_{\rm J})]}{2}\frac{1}{\pi}\frac{\Gamma}{\hbar^2(\omega_{\rm J}-\omega)^2+\Gamma^2},
\end{equation}
where we assume a small $\Gamma=4\pi k_BT Z_0/R_{\rm Q}\equiv 2\hbar D$ compared to $\hbar\omega_{\rm J}$.
The same result is obtained also from (\ref{OutputFlux}) by inserting phase correlations of classical (thermal) phase diffusion~\cite{IngoldEuro2}, $J(t)=-D\vert t\vert$.
Especially, for $T=0$ one has $J(t)=0$ and $P(E)=\delta(E)$. Then all the radiation is emitted at $\omega=\omega_{\rm J}$ with the total power $I_{\rm c}^2{\rm Re}[Z_{\rm t}(\omega_{\rm J})]/2$. In an exact classical solution also higher harmonics
and a change in the  dc-voltage across the junction  exist, but the main picture remains.

In the quantum-mechanical treatment of the EM fields, two qualitative differences appear when $T\rightarrow 0$: (i) the linewidth remains finite due to shot noise in the charge transport~\cite{Likharev,DahmScalapino,Stephen1,Lee,Rogovin,Levinson,Koch,IngoldEuro2002} and (ii)  radiation below $\omega_{\rm J}$ has a finite tail due to multi-photon emission~\cite{Hofheinz2011,Leppakangas2013}. Whereas property (i) is not captured by the leading-order perturbation theory done here
(except for the derivation of the zero-frequency shot noise, see section~\ref{sec:convergence}),
property (ii) is seen as an asymmetric broadening of the $P(E)$-function. At zero temperature and for $Z_0=Z_1$, the
$P(E)$-function has an approximative form ($E>0$)~\cite{Devoret1990,Ingold1992},
\begin{equation}\label{ZeroTemperature}
P(E)=\frac{\exp(-2\rho\gamma_{\rm E})}{\Gamma(2\rho)}\frac{1}{E}\left[ \pi\rho\frac{E}{4E_{C_{\rm J}}} \right]^{2\rho},
\end{equation}
where $\gamma_{\rm E}$ is the Euler constant, $E_{C_{\rm J}}=e^2/2C_{\rm J}$ is the junction charging energy, and $\rho=Z_0/R_Q$ is the dimensionless resistance of the transmission line.

For $E<0$ one has $P(E)=0$, i.e. no energy can be extracted from the environment.
This is obtained by using~\cite{Devoret1990,Grabert1988,Ingold1999}
$J(t)=-2\rho\left[ \ln(\omega_R \vert t\vert)+\gamma + i\frac{\pi}{2}{\rm sign}(t) \right]$. This zero temperature
long-time behaviour is a good approximation also at finite temperatures for frequencies $\omega<\omega_{\rm J}-k_{\rm B}T/\hbar$.
For a typical low-Ohmic transmission line, $\rho \ll 1$, the resulting power density is peaked at the
Josephson frequency $\omega_{\rm J}$
with the magnitude $\sim  I_C^2Z_0 \delta(\omega_{\rm J}-\omega)/2$. A finite tail extends to lower frequencies, $\omega<\omega_{\rm J}-k_{\rm B}T/\hbar$, with the form
\begin{equation}\label{eq:FluxBelow}
f_{\rm t}(\omega)\approx I_{\rm c}^2 Z_0\frac{\rho}{\hbar\omega(\omega_{\rm J}-\omega)}.
\end{equation}
This is symmetric around half the Josephson frequency, $\omega_{\rm J}/2$,
indicating that the radiation results from photons created in pairs~\cite{Leppakangas2013} whose frequencies $\omega_a$ and $\omega_b$ add up to the Josephson frequency $\omega_{\rm J}$. 
This result can also be derived also by straight linearization
of the boundary condition, which includes maximally two photon emission processes, as demonstrated in Appendix~A.
Similar results hold also for the cavity configuration, $Z_1 > Z_0$.
Especially, if the Josephson frequency matches with the sum of the frequency of two modes, strong pair production to these
modes is observed.
Numerical results for the photon-flux density for the free-space and cavity configurations are presented in figure~\ref{figure2}.

\subsection{Elastic and inelastic reflection of thermal photons, $f_{\rm th}(\omega)$}

In addition to the radiation created by inelastic Cooper-pair tunneling,
the leading-order result (\ref{FluxGeneral})
has a term proportional to the Bose factor,
which we further divide as $f_{\rm th}(\omega)=f_{0}(\omega)+f_{\rm in}(\omega)$.
The part $f_{0}(\omega)$ describes the zeroth order (elastic) reflection of photons at the junction,
\begin{equation}\label{Thermal}
f_{0}(\omega)\equiv\int_0^{\infty} d\omega '\frac{1}{2\pi}\left\langle a^{\dagger}_{0}(\omega)a_{0}(\omega')\right\rangle =\frac{1}{2\pi}\frac{1}{e^{\beta\hbar\omega}-1}.
\end{equation}
The inelastic term $f_{\rm in}(\omega)$ comes from correlators
between the zeroth and second-order operators. For the free-space configuration, $Z_0=Z_1$, we obtain (Appendix C)
\begin{equation}
f_{\rm in}(\omega)=\frac{1}{e^{\beta\hbar\omega}-1}    \frac{ I_{\rm c}^2 {\rm Re}[Z_{\rm t}(\omega)]}{2\omega} \sum_{\pm }\left[P(\pm \hbar\omega_{\rm J}-\hbar\omega) -  P(\pm\hbar\omega_{\rm J}+\hbar\omega)  \right] . \label{eq:FluxThermalInelastic}
\end{equation}

We interpret this as inelastic reflection of thermal photons,
exchanging energy with a Cooper-pair tunneling either direction.
The term proportional to $P(\pm \hbar\omega_{\rm J}-\hbar\omega)$ contributes as photon emission to the frequency
$\omega$, and the term  $\propto P(\pm \hbar\omega_{\rm J}+\hbar\omega)$ as photon absorption from this frequency.
Such processes do not contribute to the net current, and are a small correction to $f_{\rm t}(\omega)$ for the situations considered in this article.

\subsection{Convergence}\label{sec:convergence}
So far we have found out that it is the phase fluctuations across the JJ that describe Cooper-pair tunneling and simultaneous photon emission in the leading order.
To study the convergence of the perturbation expansion, we then investigate the spectrum of the phase fluctuations at the junction.
In particular, we compare the magnitude of the zeroth-order contribution with the magnitude of the leading-order contribution. For a rapidly converging perturbation expansion, the latter should be much smaller than the first. This should hold for all frequencies, since the right-hand side of the boundary condition at the junction (\ref{RCJ}), mixes all combinations of frequency terms summing up to $\omega_{\rm J}$.
This leads to the comparison
\begin{equation}
4{\rm Re}[Z_{\rm t}(\omega)]\left\langle a_0^{\dagger}a_0 + a_{0}a_0^{\dagger} \right\rangle \gg {\rm Re}[Z_{\rm f}(\omega)] \left\langle a_1^{\dagger}a_1 + a_{1}a_1^{\dagger} \right\rangle. \nonumber
\end{equation}
Here we have defined ${\rm Re}[Z_{\rm f}(\omega)]=Z_1\vert\kappa_+ +\kappa_-e^{2ik^{\rm c}_{\omega}d}\vert^{2}$,
where $\kappa_{\pm}=\left(\sqrt{Z_0/Z_1}\pm \sqrt{Z_1/Z_0} \right)/2$.
This leads us to the general condition
\begin{equation}\label{GeneralCondition}
\frac{2}{\pi}\coth\left(\frac{\beta\hbar\omega}{2}\right) \gg \frac{I_{\rm c}^2}{2\omega}{\rm Re}[Z_{\rm f}(\omega)]\sum_{\pm\pm}[P(\pm \hbar\omega_{\rm J} \pm\hbar\omega)].
\end{equation}
This is a relation for the smallness of the junction's critical current $I_{\rm c}$.
For frequencies below the cut-off frequency $\omega_c$, ${\rm Re}[Z_{\rm f}(\omega)]$  is approximately equal to ${\rm Re}[Z_{\rm t}(\omega)]$, and
in the following we will replace $Z_{\rm f}(\omega)$ by $Z_{\rm t}(\omega)$. 
We now examine the condition (\ref{GeneralCondition}) explicitly at zero frequency, at the Josephson frequency, and then finally for frequencies between these special frequencies.

Let us investigate the zero frequency limit by multiplying each side of (\ref{GeneralCondition}) by $\hbar\omega$. We get
\begin{equation}
\frac{4k_{\rm B} T}{\pi}\gg \hbar I_{\rm c}^2 {\rm Re}[Z_{\rm t}(0)][P(\hbar\omega_{\rm J})+P(-\hbar\omega_{\rm J})].
\end{equation}
Using the leading-order result for the simultaneous electric current
$I=I_{\rm c} \left\langle \sin[\phi(t)-\omega_{\rm J}t] \right\rangle=I^+-I^-$,
where $I^{\pm}(V)=(\pi\hbar I_{\rm c}^2/4e)P(\pm \hbar\omega_{\rm J})$, 
we obtain the relation
\begin{equation}\label{current}
\frac{4k_{\rm B} T}{{\rm Re} [Z_{\rm t}(0)]}\gg 4e\left[ I^+(V) + I^-(V) \right].
\end{equation}
This compares Johnson-Nyquist current noise (left-hand side) with the shot-noise coming from the charge transport (right-hand side), the noise considered also in Refs.~\cite{Likharev,DahmScalapino,Stephen1,Lee,Rogovin,Levinson,Koch,IngoldEuro2002}.
For an Ohmic impedance ($Z_0=Z_1$) this implies
\begin{equation}\label{condition1}
k_BT\gg \pi^2\rho^2\frac{E_{\rm J}}{2eV} E_{\rm J}.
\end{equation}

The second condition is calculated at the Josephson frequency $\omega=\omega_{\rm J}$.
The contribution from $\omega_{\rm J}$ would act back to the low-frequency spectrum in the next perturbation round
and would affect, for example, to a possible shift in the average voltage across the junction.
Analysis at the Josephson frequency gives us
\begin{equation}
4k_BT\gg \hbar I_{\rm c}^2\frac{{\rm Re}[Z_{\rm t}(\omega_{\rm J})]}{2eV}\frac{R_Q}{2\pi {\rm Re}[Z_{\rm t}(0)]}.
\end{equation}
To derive this, we have used the approximation $P(0)\approx (1/\Gamma\pi)$. We assume now, that the Josephson frequency does not match with a resonance
frequency in the cavity (but it still can match a sum of two). We have then ${\rm Re}[Z_{\rm t}(\omega_{\rm J})]\approx {\rm Re}[Z_{\rm t}(0)]$, and we get the condition
\begin{equation}\label{thermal}
4k_BT\gg \frac{E_{\rm J}}{2eV} E_{\rm J}.
\end{equation}
For an Ohmic impedance this is usually slightly more strict as (\ref{condition1}), as typically $\rho\sim 1/20>1/2\pi$. It is also independent of $Z_0$. For the Ohmic case
this can be then converted to a demand that thermal dephasing has to be faster than inelastic Cooper-pair tunneling,
since $D\gg I(V)/2e$ is equivalent with $2k_{\rm B}T\gg E_{\rm J}^2/2eV$ [under the approximation $P(\hbar\omega)\approx 2\rho/\hbar\omega$]. However, if the Josephson frequency is exactly at the resonance, we get
\begin{equation}\label{thermal}
4k_BT\gg \frac{E_{\rm J}}{2eV} E_{\rm J}Q_0^2.
\end{equation}
Here we have used the result for the $Q$-factors of the resonance modes $Q_n=(2n+1)\pi Z_1/4Z_0$ ($n=0,1,2,\ldots $).

For the analysis at the middle frequencies $k_{\rm B}T/\hbar< \omega< \omega_{\rm J}-k_{\rm B}T/\hbar $ we consider the approximation $P(\hbar\omega)\approx 2Z_{\rm t}(\omega)/R_Q\hbar\omega$, which gives
\begin{equation}\label{voltage}
1 \gg \frac{eI_{\rm c} Z_{\rm t}(\omega)}{\hbar\omega}\frac{eI_{\rm c} Z_{\rm t}(\omega_{\rm J}-\omega)}{\hbar(\omega_{\rm J}-\omega)}.
\end{equation}
For a resonant environment this sets a limit between the critical current and the sharpness ($Q$-factor) of the mode.
One gets then the condition  $I_{\rm c}Z_1 Q_0\ll V$.
For an Ohmic impedance we get the condition $I_{\rm c}Z_0\ll V$,
a known convergence condition for the higher orders of $P(E)$-theory obtained in reference~\cite{IngoldEuro2002}.



\section{Emission characteristics II: Second-order coherence}\label{SecondOrderCoherence}
To study statistics of the emitted photons in more detail, we investigate
the second-order coherence $G^{(2)}(\tau)$ for the output radiation, i.e. the probability to detect a pair of photons with time interval $\tau$.
The possibility for multi-photon emission implies bunching of the outgoing photons, meaning an increased probability of detecting 
photon pairs simultaneously. In this section,
we consider our results for $G^{(2)}(\tau)$, obtained by including the leading contributions up to the fourth order in
the critical current $I_{\rm c}$.

\subsection{Photon coherences}

We start with the first-order coherence, $G^{(1)}(\tau)$, for a continuous-mode field defined as~\cite{QOBook}
\begin{eqnarray}
G^{(1)}(\tau)\equiv\frac{\hbar Z_0}{4\pi}\int_0^{\infty}d\omega d\omega'\sqrt{\omega\omega'}\left\langle a_{\rm out}^\dagger(\omega) a_{\rm out}(\omega')  \right\rangle e^{i\omega\tau}. \nonumber
\end{eqnarray}
Here we use the notation $\int_0^{\infty}d\omega d\omega'\equiv \int_0^{\infty}d\omega \int_0^{\infty} d\omega'$.
Similarly as before, we estimate this up to second order in $I_{\rm c}$.
We can relate this to the photon-flux density, equation (\ref{FluxGeneral}), and obtain
\begin{eqnarray}
G^{(1)}(\tau)=\frac{\hbar Z_0}{2}\int_0^{\infty}d\omega \omega \left[ f_{\rm t}(\omega)+f_{\rm th}(\omega) \right] e^{i\omega\tau}. \nonumber
\end{eqnarray}
In the following, we are interested in the contribution due to inelastic Cooper-pair tunneling, $f_{\rm t}(\omega)$,
\begin{equation}\label{BounchingFirst}
G_{\rm t}^{(1)}(\tau)=\frac{I_C^2Z_0}{4}\int_0^{\infty}\hbar d\omega e^{i\omega\tau}{\rm Re}[Z_{\rm t}(\omega)] P(\hbar\omega_{\rm J}-\hbar\omega).
\end{equation}
Here, we have neglected the vanishing contribution due to backward Cooper-pair tunneling, $\propto P(-\hbar\omega_{\rm J}-\hbar\omega) $.

The second-order coherence, gives information of correlations between the emitted photons.
This is defined for a continuous-mode field as~\cite{QOBook}
\begin{eqnarray}\label{eq:SecondCoherence}
&&G^{(2)}(\tau)\equiv \left( \frac{\hbar Z_0}{4\pi} \right)^2\int_0^{\infty}d\omega d\omega' d\omega'' d\omega''' \sqrt{\omega\omega'\omega''\omega'''}   \nonumber \\
&\times & e^{i\tau(\omega'-\omega'')}\left\langle a_{\rm out}^{\dagger}(\omega) a_{\rm out}^{\dagger}(\omega') a_{\rm out}(\omega'') a_{\rm out}(\omega''') \right\rangle.
\end{eqnarray}
The leading-order contribution for (\ref{eq:SecondCoherence})  comes again from the
second order in $I_{\rm c}$, which describes the effect of single-Cooper-pair tunneling. To obtain analytical results
we calculate $G^{(2)}_{\rm t}(\tau)$ for the JJ connected directly to the free space, $Z_0=Z_1$, at very low temperatures (for a more general expression see Appendix D).
After a straightforward calculation we get
\begin{equation}\label{BounchingGeneral}
G^{(2)}_{\rm t}(\tau) = \left(\frac{I_C^2Z_0^2}{4}\right)^2 \left( \frac{1}{\pi E_{\rm J}} \right)^2 \int_0^{\infty}\hbar d s \left( \frac{2\hbar}{\tau} \right)^2\sin^2\left( \frac{\tau s}{2} \right)   P[\hbar(\omega_{\rm J}-s)].
\end{equation}
Here, we have neglected terms proportional to the Bose factor, i.e.~$\propto f_{0}(\omega)$. In the following we use this result
to study photon bunching in the output radiation.


\begin{figure}[bt]
\includegraphics[width=0.8\linewidth]{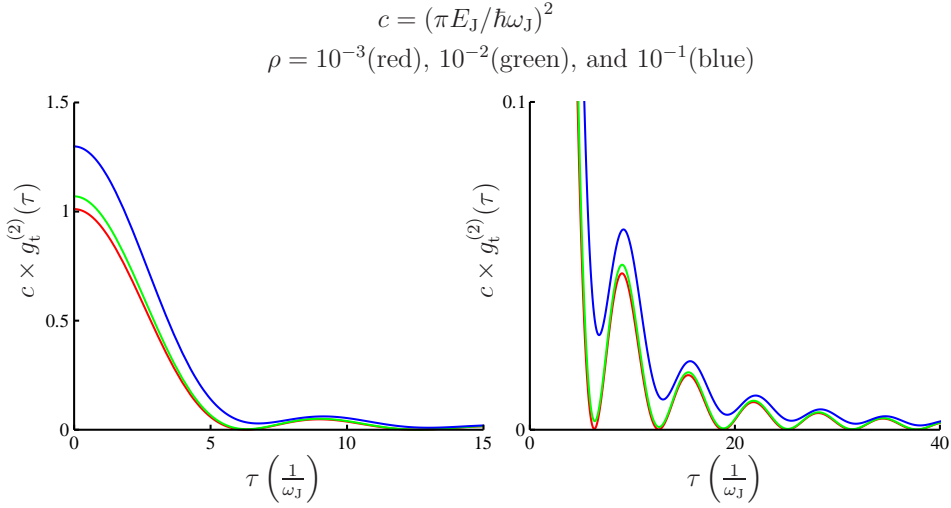}
\caption{The normalized second-order coherence for a JJ connected directly to the free space,
as estimated from~(\ref{BounchingFirst}) and~(\ref{BounchingGeneral}) using the $P(E)$-function~(\ref{ZeroTemperature}). We use here
$\rho=10^{-3}$, $10^{-2}$, and $10^{-1}$, from the bottom to top, and we have chosen $E_C=\hbar\omega_{\rm J}$.
We see that the bunching, $g^{(2)}_{\rm t}(0)$, is close to the estimate~(\ref{main:bunching})
when $\rho\ll 1$.
The second-order coherence decays due to finite bandwidth,
and has analogous form as the intensity pattern in single-slit diffraction.
}\label{figureg2}
\end{figure}

\subsection{Bunching}
An important quantity describing photon emission is the relation between the first and second-order coherences,
\begin{equation}
g^{(2)}(\tau)=\frac{G^{(2)}(\tau)}{[G^{(1)}(0)]^2}.
\end{equation}
This basically compares  probabilities for single and two photon detection. If $g^{(2)}(0)<1$, the field is called antibunched, and if $g^{(2)}(0)>1$ the field is bunched. For a Poissonian process the result is $g^{(2)}(0)=1$, while for thermal radiation $g^{(2)}(0)=2$.
Arbitrarily high bunching is possible also classically whereas antibunching is a pure sign of nonclassicality~\cite{QOBook}.

With the analytical values of the first and the second-order coherences
we can immediately get an estimate for the bunching in the free-space configuration ($Z_0=Z_1$).
We consider a typical transmission line ($\rho\ll 1$, $\omega_{\rm J}\ll \omega_{\rm c}$) and solution (\ref{ZeroTemperature}), and obtain (Appendix D)
\begin{equation}\label{main:bunching}
g^{(2)}_{\rm t}(0)=\frac{G^{(2)}_{\rm t}(0)}{[G^{(1)}_{\rm t}(0)]^2}\approx \left( \frac{\hbar\omega_{\rm J}}{\pi E_{\rm J}} \right)^2.
\end{equation}
This can be made arbitrary large by decreasing the critical current, i.e., the emitted power.
This property is typical for a pair production of photons. Notably is that result~(\ref{main:bunching}) is independent of $\rho$,
even though the power is proportional to $\rho$. In figure~\ref{figureg2}, we visualize the time dependence of $g^{(2)}_{\rm t}(\tau)$.

As Cooper-pair tunneling is also accompanied by an emission of low-energy photons, describing a simultaneous change in the voltage across the
junction, it is more clear for the interpretation of the results not to include frequencies in the neighborhood of $\omega=0$ or $\omega=\omega_{\rm J}$. We consider then a small interval $\Delta\omega$ of frequencies around half the Josephson frequency $\omega_J/2$, i.e.~$\omega_{\rm J}/2-\Delta\omega/2<\omega<\omega_{\rm J}/2+\Delta\omega/2$, which in an experiment corresponds to a filtering of the
output radiation~\cite{Silva}.
One obtains for the corresponding second-order coherence (Appendix D)
\begin{eqnarray}
G^{(2)}_{\rm t}(0) \approx  \left(\frac{I_{\rm c}^2Z_0^2}{4}\right)^2  \left(\frac{\hbar \Delta\omega}{\pi E_{\rm J}}\right)^2 . \nonumber
\end{eqnarray}
The related first-order coherence, within the same approximation, is
$ G^{(1)}(0) \approx  \rho I_{\rm c}^2  Z_0^2 \Delta\omega/ \omega_{\rm J}$.
Therefore, we obtain the bunching, if measured in a small frequency interval around $\omega_{\rm J}/2$,
\begin{equation}\label{eq:BunchingSmallInterval}
g^{(2)}_{\rm t}(0)\approx \left( \frac{1}{4\rho}\right)^2 \left( \frac{\hbar\omega_{\rm J}}{\pi E_{\rm J}} \right).
\end{equation}
This is proportional to $1/\rho^2$, and for the considered case of small $\rho$, is much larger than result (\ref{main:bunching}) for the complete output field.

\begin{figure}[bt]
\includegraphics[width=0.85\linewidth]{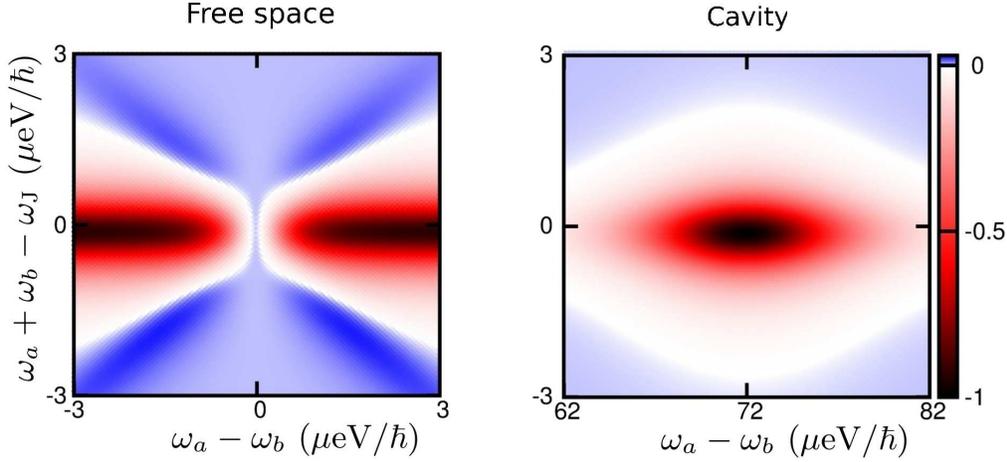}
\caption{We plot here the violation of the Cauchy-Schwarz inequality (\ref{Test2}), ${\cal N}=  P\left[\hbar(\omega_d-2\omega_a)\right] P\left[\hbar(\omega_d-2\omega_b)\right]-P\left[\hbar(\omega_d-\omega_a-\omega_b)\right]^2$, 
multiplied by the photon-flux densities at the two frequencies, i.e.~${\cal N}\times f_{\rm t}(\omega_a) f_{\rm t}(\omega_b)$.
The colour scale is normalised to the maximal value, which is  $2\times 10^{5}$ higher for the cavity configuration (right) compared to the JJ connected directly to the free space (left).
Negative values are a sign of nonclassicality and the parameters correspond to figure~\ref{figure2}.
The violation occurs around the condition of photon pair production $\omega_a+\omega_b= \omega_{\rm J}=140$~$\mu$eV$/\hbar$.
For the cavity configuration the observed nonclassicality is enhanced when the two frequencies match the two lowest modes of the cavity, maximising the photon pair production. }
\label{figure4}
\end{figure}

If we consider detection in a small frequency interval completely above half the Josephson frequency $\omega_{\rm J}/2$, we obtain the leading-order result
$g^{(2)}_{\rm t}(0)\approx 0$ (Appendix D), and exactly zero at the zero temperature.
This is because the photon production at these frequencies occurs through
multi-photon processes, and emission of two (or more) photons above $\omega_{\rm J}/2$ is not possible from a single-Cooper-pair
process.
However, the result $g^{(2)}_{\rm t}(0)=0$ does not imply that the field is antibunched,
since contributions from higher orders is neglected.
The next-order contribution for $G^{(2)}_{\rm t}(0)$ comes from the fourth order, which has a special meaning
as this is also the leading order of $\vert G^{(1)}_{\rm t}(0)\vert^2$. This order is also the first one to describe
photon emission from two Cooper-pair tunnelings.
Analytical results can be obtained for the Ohmic environment in the considered case $\rho\ll 1$.
The most important contribution becomes from a term describing two photon emission
processes due to two (correlated) Cooper-pair tunnelings,
$\left\langle a_1^{\dagger}a_1^{\dagger}a_1a_1\right\rangle$.
For small $\rho$ and approximation $J(t)=-D\vert t\vert -i\pi\rho{\rm Sgn}(t)$~\cite{IngoldEuro2}, we get through a lengthy analytical calculation a contribution
$
g^{(2)}_{\rm t}(0)=2-\tilde B,
$
where $\tilde B\in (1,2)$. At the Josephson frequency and for
a bandwidth larger than thermal dephasing $D$,
we get $\tilde B\approx 1$. Therefore the photon characteristics nearby
this frequency is close to a Poissonian. Well below $\omega_{\rm J}$
the calculation gives $\tilde B<<1$.


\section{Nonclassicality}\label{Nonclassicality}
The electromagnetic field is nonclassical if it cannot be described by the classical theory of electromagnetism.
One example is a quadrature squeezed state of single mode, for which the width of the Wigner quasiprobability distribution in one of the quadratures is smaller than the width of a coherent state, i.e.~the quantum description of a classical coherent signal.~\cite{Milburn}
The quadrature squeezing is measured through amplitude auto- and cross-correlations, which are phase-sensitive quantities.
In our system, the Josephson junction is driven by a dc voltage, which suffers from both thermal and transport noise.
As we will see, this leads to a rather short phase coherence time and no steady-state squeezing.
There exists however a number of other relations, that are satisfied by a classical field, but can be violated by a general quantum mechanical field.~\cite{Nonclassical1}
These are useful in our system, if they are immune to dephasing.
An example of such a nonclassicality test is the Cauchy-Schwarz inequality for intensity auto- and cross-correlations, that is known to be
violated maximally for a field created through parametric down conversion~\cite{Milburn}.

In this section, we first address the question of quadrature squeezing in the output radiation,
and then go on to derive a Cauchy-Schwarz inequality for photon-flux correlations in the leading-order approximation,
which we find to be an optimal way of detecting nonclassicality in the considered system.

\subsection{Quadrature squeezing and dephasing}
The pair production of photons implies quadrature squeezing,~\cite{Milburn} which is
characterized by correlators of type $\left\langle a_{\rm out}(\omega)a_{\rm out}(\omega')\right\rangle$.
The result (\ref{eq:Amp2}), however, means that such nonclassical correlations do not exist on average, due to
dephasing of the phase difference across the junction.
This can be qualitatively visualized as a diffusion
of the angle of quadrature squeezing. 
The situation is analogous to a parametric down conversion with an nonideal drive~\cite{QOBook}.

To investigate how long it takes for the squeezing angle to be randomized, if one would know its value (or distribution) at
time $t=0$, we consider the phase-coherence function,
\begin{equation}
\left\langle e^{i\phi_0(t)}e^{-i\phi_0(0)}\right\rangle=e^{J(t)}.
\end{equation}
In the long-time limit and for finite temperatures its behaviour is defined by the zero-frequency impedance $Z_0$, via
$J(t)\sim -D\vert t\vert$, where $D=2\pi k_{\rm B}T\rho/\hbar$.
Therefore, we identify $D$ as the dephasing rate of quadrature squeezing.
For typical values for the low-frequency impedance $Z_0=50$~$\Omega$ and $T=20$~mK, one has $1/D\approx 8$~ns.
Such dephasing times are therefore a very relevant property of a voltage-driven system,
and a challenge for a  measurement  of phase-dependent system properties.

\subsection{ Cauchy-Schwarz inequality for intensity cross-correlations}
A nonclassicality test that is not affected by phase fluctuations must be of higher order in the
operators $a^{(\dagger)}_{\rm out}$. The logical thing to do is to add two more operators to the ensemble average that
characterizes squeezing, $\langle a_{\rm out}a_{\rm out} \rangle$.
Basically we have two possibilities to consider, the second-order coherence, of type
$\langle a^{\dagger}_{\rm out}a^{\dagger}_{\rm out}    a_{\rm out}a_{\rm out} \rangle$,
or the intensity correlator, of type $\langle a^{\dagger}_{\rm out} a_{\rm out} a^{\dagger}_{\rm out} a_{\rm out} \rangle$.
The second-order coherence was considered in Section~\ref{SecondOrderCoherence}, and was found to reveal a high degree of bunching,
as a result of photon pair production. However, only antibunching would be a proof of nonclassicality. Therefore, we will now
investigate a nonclassicality condition based on intensity cross-correlations.
In the countable-mode case, a suitable Cauchy-Schwarz inequality is of the form~\cite{Nonclassical1}
\begin{equation}
\left\vert \left \langle a_1^{\dagger}a_1a_2^{\dagger}a_2 \right \rangle\right \vert^2 \leq \left \langle (a_1^{\dagger})^2a_1^2\right \rangle  \left \langle (a_2^{\dagger})^2a_2^2 \right \rangle .
\end{equation}
In the following, we apply this condition to the considered continous-mode case.

In the case of continuum of modes, we practically estimate $G^{(2)}(0)$ over a small frequency range $\Delta\omega$ around $\omega_1$ or $\omega_2$, and similarly for the corresponding cross correlator.
Through a straightforward calculation we obtain for the considered auto-correlator (Appendix D)
\begin{eqnarray}\label{eq:Auto}
&&\int_{\omega_a-\Delta\omega/2}^{\omega_a+\Delta\omega/2}d\omega_1 d\omega_2 d\omega_3 d\omega_4  \left \langle a^{\dagger}_1a^{\dagger}_2a_3a_4\right \rangle   \\
&&=\frac{2\pi I_{\rm c}^2 \Delta\omega^3}{\omega_a^2 R_{\rm Q}}P\left[ \hbar(\omega_{\rm J}-2\omega_a) \right]\left[{\rm Re}\left[ Z_{\rm t}(\omega_a) \right]\right]^2 +{\cal O}(\Delta\omega^4).\nonumber
\end{eqnarray}
To keep the notation short we mark here $a_i\equiv a(\omega_i)$.
$G^{(2)}(0)$ at $\omega_a$ is calculated up to the second-order in $I_{\rm c}$ and similarly for the contribution at $\omega_b$.

On the other hand, the intensity cross-correlations between the two frequencies have the form
(when $\vert \omega_a-\omega_b\vert  > \Delta\omega$)
\begin{eqnarray}
&&\int_{\omega_a-\Delta\omega/2}^{\omega_a+\Delta\omega/2}d\omega_1 d\omega_2 \int_{\omega_b-\Delta\omega/2}^{\omega_b+\Delta\omega/2}d\omega_3 d\omega_4 \left \langle a^{\dagger}_1a_2a^{\dagger}_3a_4 \right \rangle \label{eq:Intensity} \\
&&  =    \frac{2\pi I_{\rm c}^2 \Delta\omega^3}{\omega_a\omega_b R_{\rm Q}}P\left[ \hbar(\omega_{\rm J}-\omega_a-\omega_b) \right] {\rm Re}\left[ Z_{\rm t}(\omega_a) \right] {\rm Re}\left[ Z_{\rm t}(\omega_b) \right] + {\cal O}(\Delta\omega^4) \nonumber
\end{eqnarray}
With the results (\ref{eq:Auto}-\ref{eq:Intensity}) we get then the
Cauchy-Schwarz inequality (in the limit $\Delta\omega\rightarrow 0$ and calculated up to second-order in $I_{\rm c}$)
\begin{equation}\label{Test2}
P\left[\hbar(\omega_{\rm J}-\omega_a-\omega_b)\right]^2\leq P\left[\hbar(\omega_{\rm J}-2\omega_a)\right] P\left[\hbar(\omega_{\rm J}-2\omega_b)\right].
\end{equation}
This result is valid for both the free-space and the cavity configuration.

The inequality (\ref{Test2}) is defined only via the $P(E)$-function (\ref{PEDefinition}).
The left-hand side of (\ref{Test2}) has a maximum when $\omega_a+\omega_b=\omega_{\rm J}$, i.e.~when the argument
goes to zero.
If at the same time $\vert \omega_a-\omega_b\vert \gg k_{\rm B} T$,
the right-hand side is close to zero, as one of the $P(E)$-functions has a large negative argument compared to the temperature.
In this case the inequality becomes violated, which we visualize in figure~\ref{figure4}.
The violation is due to nonclassical photon pair production.
Generally at $\omega_a=\omega_b$ the nonclassicality cannot be tested with this inequality since the two sides are equal by definition.
The use of a resonant environment ($Z_1\gg Z_0$) does not change the violation of the inequality~(\ref{Test2}) qualitatively. 
However, it significantly increase the photon emission rate at certain frequencies, which facilitates experimental detection~\cite{Leppakangas2013}.
As the contribution from thermal radiation is neglected here,
the tested frequencies $\omega_{a(b)}$ should be well above $k_{\rm B}T/\hbar$.
Also, in an experiment a detection over a finite bandwidth is used, whose effect should be carefully analyzed.

\section{Conclusions and outlook}

In conclusion, we have derived a continuous-mode solution for microwave radiation in a transmission line with a dc-voltage bias and which is
terminated by a small Josephson junction.
This is done by a perturbative treatment of the boundary condition
that describes Cooper-pair tunneling across the Josephson junction.
We showed that the method reproduces
the previously derived expression for the created photon-flux density, obtained by applying the $P(E)$-theory.
We extended this first by determining the corresponding second-order coherence.
We found that the emitted microwave field has a high degree of bunching due to photon pair production at frequencies below the Josephson
frequency. We then addressed the question of nonclassicality of the emitted radiation
in this region and showed that the photon pair production violates the classical Cauchy-Schwarz inequality for
intensity cross-correlations.

The established method opens a possibility for further detailed study of the radiation characteristics in this system.
For example, calculations in higher order access the question of the effect of correlations between consecutive tunneling Cooper pairs.
For the considered case of low-Ohmic environment, we obtained bunching in the output radiation.
On the other hand, when the transmission-line impedance is increased beyond the resistance quantum,
antibunching of the Cooper-pair tunneling is expected, due to Coulomb blockade.~\cite{Ingold1992}
In this regime, the output photons should also be antibunched.
Also, summation to all orders can be
feasible, if known summation methods for this type of perturbation expansions work.
Overall, this system is very rich in physics, covering the limit of dynamical Coulomb blockade at low impedances,
to Coulomb blockade in the high-impedance limit. The question of the detailed form and properties of the related output radiation makes this system very interesting for future works.
This is motivated also by the technical development towards simultaneous measurements of both microwaves and electrical currents.

\appendix

\section*{Appendix A: Linearization of the boundary condition at the junction}
A straightforward way to solve for the out-field is to linearize the boundary condition~(\ref{RCJ}) and Fourier transform the problem.
The silent assumption is the small fluctuations of the phase difference, $\phi(t)$. This is actually not usually the case,
since the (zeroth order) phase difference performs a quantum Brownian motion in time~\cite{Ingold1992,Grabert1988}.
However, the linearization turns out to give correct results for frequencies $\omega\sim\omega_{\rm J}/2$ ($\rho\ll 1$),
where such fluctuations stay small. This is consistent with pair production of photons in this frequency range.

In the following we consider linearization in the case of the free-space configuration, $Z_0=Z_1$,
and take the limit $C_{\rm J}\rightarrow 0$.
To do this properly, we rewrite the right-hand side of (\ref{RCJ}) using the identity
\begin{eqnarray}
\sin[\phi(t)-\omega_{\rm J} t]=-\cos \phi(t)\sin \omega_{\rm J} t+\sin \phi(t)\cos \omega_{\rm J} t.\nonumber
\end{eqnarray}
Expanding the right-hand side of this up to linear order in $\phi(t)$, and Fourier transforming, we get for this
\begin{eqnarray}
\frac{\pi}{i}[\delta(\omega-\omega_{\rm J}) -\delta(\omega+\omega_{ \rm J} )]+\sum_{\pm}\frac{\pi^{3/2}}{\Phi_0}\sqrt{\hbar Z_0}\frac{1}{\sqrt{\vert \omega\pm\omega_{\rm J} \vert}}[a_{\rm in}(\omega\pm\omega_{\rm J} )+a_{\rm out}(\omega\pm\omega_{\rm J} )].\nonumber
\end{eqnarray}
Here, for simplicity, we have introduced negative frequencies as $a(-\omega)\equiv a^{\dagger}(\omega)$.
The first two terms represent radiation at the Josephson frequency $\omega_{\rm J}$, while the other terms
describe mixing of this with additional photonic process where, for example, $\omega_{\rm J}$ is splitted into
two frequencies.

We continue by solving the corresponding boundary condition,
\begin{eqnarray}
&&a_{\rm out}(\omega)+i{\rm Sgn}(\omega)\frac{Z_0I_{\rm c} \pi}{\Phi_0}\sqrt{\frac{1}{\omega}}\sum_{\pm}\frac{a_{\rm out}(\omega\pm\omega_{\rm J} )}{\sqrt{\vert \omega\pm\omega_{\rm J} \vert}}=\nonumber \\
&&a_{\rm in}(\omega)-i{\rm Sgn}(\omega)\frac{Z_0I_{\rm c} \pi}{\Phi_0}\sqrt{\frac{1}{\omega}}\sum_{\pm}\frac{a_{\rm in}(\omega\pm\omega_{\rm J} )}{\sqrt{\vert \omega\pm\omega_{\rm J} \vert}}.\nonumber
\end{eqnarray}
This can be done by writing the equation into the matrix form
\begin{equation}\label{MatrixEquation}
M_{\rm out}(\delta\omega)A_{\rm out}(\delta\omega)=M_{\rm in}(\delta\omega)A_{\rm in}(\delta\omega).
\end{equation}
Here the frequency vector $A$ is constructed as~\cite{Johansson2010}
\begin{eqnarray}
A^T=\{a[-N\omega_{\rm J}+\delta\omega], \ldots ,a[N\omega_{\rm J}+\delta\omega] \},\nonumber
\end{eqnarray}
where $\vert\delta\omega\vert<\omega_{\rm J}$. This form is possible since the boundary condition mixes only frequencies differing by $\omega_{\rm J}$.
We have also introduced a cut-off $\Omega=N\omega_{\rm J}$. 
The matrices $M$ have diagonals 1 and first nondiagonals (in the $n$:th row)
$d^{\pm}[(-N-1+n)\omega_{\rm J}+\delta\omega]$, where the
plus sign corresponds to the term $M_{n,n+1}$
and we have
\begin{eqnarray}
d^{\pm}_{\rm out}(\omega)={\rm Sgn}(\omega)i\pi\frac{I_{\rm c}}{\Phi_0}Z_0\frac{1}{\sqrt{\vert \omega \vert\vert \omega\pm\omega_{\rm J} \vert}}, \nonumber
\end{eqnarray}
and $d_{\rm in}=-d_{\rm out}$. Equation (\ref{MatrixEquation})
has to be solved generally numerically.
For small $d$ approximative analytical solution can be sought with the ansats
\begin{equation}\label{Ansats}
a_{\rm out}(\omega)=a_{\rm in}(\omega)+S^{+}(\omega)a_{\rm in}(\omega+\omega_{\rm J} )+S^{-}(\omega)a_{\rm in}(\omega-\omega_{\rm J}).\nonumber
\end{equation}
We then find a solution in the lowest order for $I_{\rm c}$
\begin{equation}\label{AnsatsSolution}
S^{\pm}(\omega)=-2d^{\pm}(\omega).
\end{equation}
We can now estimate the photon flux density below the Josephson frequency. Using~(\ref{AnsatsSolution}) one gets
for $\omega<\omega_{\rm J}$ and at $T=0$
\begin{equation}\label{LinearizedAnalytic}
f_{\rm t}(\omega) \frac{1}{2\pi}\vert S^-(\omega)\vert^2=\frac{\rho I_{\rm c}^2Z_0}{\hbar\omega(\omega_{\rm J}-\omega)}. \nonumber
\end{equation}
This result consistent with~(\ref{eq:FluxBelow}).



\section*{Appendix B: Perturbative input-output approach}
\subsection*{Zeroth-order solution}\label{LinearSolutionApp}
After Fourier transformation of boundary conditions~(\ref{eq:BoundaryLine1}-\ref{eq:BoundaryLine2}) we get
\begin{eqnarray}
a_{\rm out}^{\rm c}(\omega)  &=& \kappa_-  a_{\rm in}^{\rm f}(\omega)  e^{-2ik_{\omega}^{\rm c}d} + \kappa_+ a_{\rm out}^{\rm f}(\omega)  \label{CavityBoundary1}, \\
a_{\rm in}^{\rm c}(\omega)  &=&\kappa_+  a_{\rm in}^{\rm f}(\omega)  + \kappa_- a_{\rm out}^{\rm f}(\omega) e^{2 ik_{\omega}^{\rm c}d}   \label{CavityBoundary2}.
\end{eqnarray}
Here $2\kappa_{\pm}=\sqrt{Z_0/Z_1} \pm \sqrt{Z_1/Z_0}$.
Considering the zeroth order solution,  $E_{\rm J}=0$, the
Fourier transformation ($\omega>0$) of
the boundary condition at the junction, equation~ (\ref{RCJ}), gives
\begin{eqnarray}
{\cal C}(\omega)a_{\rm out}^{\rm c}(\omega) - {\cal C}^*(\omega)a_{\rm in}^{\rm c}(\omega)=0.
\end{eqnarray}
The solution is then
$a_{\rm out}^{\rm c}(\omega)= e^{i\theta(\omega)}a_{\rm in}^{\rm c}(\omega)$.
Combining this with equations~(\ref{CavityBoundary1}-\ref{CavityBoundary2}) we obtain (\ref{ZerothSolutionMain}-\ref{PhaseDifference0})

\subsection*{Higher-order solution}
For higher orders we have no input field from the free space and the boundary condition at $x=d$ gets the form ($n\geq 1$)
\begin{eqnarray}
b_n(\omega)&=&\kappa_+a_n(\omega), \\
c_n(\omega)&=&\kappa_-a_n(\omega)e^{2 ik_{\omega}^{\rm c}d} .
\end{eqnarray}
Here the output (input) cavity field in the $n$:th order is labelled as $b_n$ ($c_n$).
We rewrite the boundary condition at the junction as
\begin{eqnarray}
&&b_n(x=0)=c_n(x=0)e^{i\theta(\omega)}+\\
&&+ i I_{\rm c}\frac{ \sqrt{ Z_1/\hbar\omega\pi } }{{\cal C}(\omega)  }\int_{-\infty}^{\infty}e^{i\omega t}dt\{\sin\left[ \phi(t)-\omega_{\rm J} t \right]\}_{n-1}. \nonumber
\end{eqnarray}
Here the formal operation $\{ \cdot \}_n$ picks out $n$:th order contribution. It follows then
\begin{equation}
a_n(\omega)=i \sqrt{\frac{Z_1}{\hbar\omega\pi}} \frac{A(\omega)}{{\cal C}^*(\omega)}I_{\rm c}\int_{-\infty}^{\infty}e^{i\omega t}dt\{\sin\left[ \phi(t)-\omega_{\rm J} t \right]\}_{n-1}.
\end{equation}
In the leading order, we can only include zeroth-order phase difference in the Taylor expansion of operators
$e^{\pm i\phi_0+\xi\phi_1+\xi^2\ldots}$, and we immedately obtain (\ref{SolutionOut1}).
Generally, one can solve for the phase difference at the junction in the $n$:th order,
\begin{eqnarray}\label{GeneralPhase}
\phi_n(t)&=&\int_{0}^{\infty}\frac{d\omega}{\sqrt{\omega}} B(\omega)a_n(\omega) e^{-i\omega t}+{\rm H.c.},\\
B(\omega)&=& \frac{\sqrt{\hbar Z_1\pi}}{\Phi_0} \left(\kappa_+ + \kappa_-e^{2ik_\omega^{\rm c} d}\right). \nonumber
\end{eqnarray}
This is self consistent, since the same order result for $a_n(\omega)$ depends only on the previous-order phase-differences.

The phase difference in the leading order has a central role when constructing the general solution.
We get for this
\begin{eqnarray}\label{PhaseLeadingOrder}
\phi_1(t)&=&  \frac{ I_{\rm c} Z_1 }{\Phi_0} \int_{-\infty}^{\infty}dt'  \sin\left[ \phi_0(t')-\omega_{\rm J} t' \right] \tilde C (t-t') \\
\tilde C (t-t')&=&   i\int_{0}^{\infty}\frac{d\omega}{\omega}       \bar A(\omega) \left(\kappa_+ + \kappa_-e^{2ik_\omega^{\rm c} d}\right)      e^{-i\omega (t-t')} +{\rm H.c.}  
\end{eqnarray}
We calculate the explicit form  for $Z_0=Z_1$,
\begin{eqnarray}\label{PhaseDifference1}
\tilde C (t-t')= \pi \left[ {\rm Sgn}(t-t')\left(1-e^{-\omega_c\vert t-t'\vert}\right)+e^{-\omega_c\vert t-t'\vert} \right].
\end{eqnarray}

\subsection*{Treatment of $\sin[\phi(t)-\omega_{\rm J}t]$ to second order}
We aim to expand the term $\xi I_{\rm c}\sin[\phi(t)-\omega_{\rm J} t]$ to second order in $\xi$.
We have formally $\phi(t)=\phi_0(t)+\xi\phi_1(t)+\ldots$,
and we need to properly expand the functions
\begin{equation}
e^{\pm i[\phi_0(t)+\xi\phi_1(t)+\xi^2\phi_2(t)+\ldots -\omega_{\rm J} t]},\nonumber
\end{equation}
to first order in $\xi$. We know that up to first order in $\xi$ we can include only operators $\phi_0(t)$
and $\xi\phi_1(t)$ in the Taylor expansion.
Therefore we can put $\phi(t)=\phi_0(t)+\xi\phi_1(t)$.
We now define an operator $z(t)$ through the relation
$\phi_1(t)=[\phi_0(t),z(t)]$. Through a direct calculation
of the Taylor expansion one gets for the first-order contribution
\begin{equation}
\left\{ e^{\pm i[ \phi_0(t)+i\xi\phi_1(t)]} \right\}_1=\xi[e^{\pm i\phi_0(t)},z(t)].\nonumber
\end{equation}

To solve $z$ we first evaluate commutator $C(t-t')=[\phi_0(t),\phi_0(t')]$ . Generally
\begin{equation}\label{eq:CommutatorCorrelator}
C(t-t')=-4i\int_{0}^{\infty}d\omega\frac{\sin\omega(t-t') }{\omega}\frac{{\rm Re}[Z_{\rm t}(\omega)]}{R_{\rm Q}}.
\end{equation}
We get for the special case $Z_0=Z_1$,
\begin{equation}
C(t-t')=-i\frac{2\pi Z_0}{R_{\rm Q}}{\rm Sgn}(t-t')\left( 1-e^{-\omega_c\vert t-t'\vert} \right).\nonumber
\end{equation}
We then investigate the commutator
\begin{eqnarray}
\left[ \phi_0(t),\cos[\phi_0(t')-\omega_{\rm J} t'] \right] =-C(t-t')\sin[\phi_0(t')-\omega_{\rm J} t'].\nonumber
\end{eqnarray}
Here we have used that
$[\phi_0(t),e^{\pm i\phi_0(t')}]=\pm i C(t-t')e^{\pm i\phi_0(t')}$.
Comparing this with solution (\ref{GeneralPhase}) we deduce
\begin{equation}
z(t)=-\frac{I_{\rm c}Z_1}{\Phi_0}\int_{-\infty}^{\infty}dt' \cos[ \phi_0(t')-\omega_{\rm J} t' ]\frac{\tilde C(t-t')}{C(t-t')}.\nonumber
\end{equation}
For the open-space configuration ($Z_0=Z_1$) we get then
\begin{eqnarray}
\frac{\tilde C(t)}{C(t)}=i\frac{R_{\rm Q}}{2Z_0}\left[ 1+\frac{{\rm Sgn}(t)}{e^{\omega_c\vert t\vert}-1} \right],\label{App:Correlator}
\end{eqnarray}
which leads to (\ref{SolutionOut2}). This expansion (with methods shown in Appendix C) can be also used to rederive
the $P(E)$-theory net current across the JJ, used in section~\ref{sec:convergence}.

\subsection*{Commutation relations for the out field}
We express the zeroth-order solution in the form
\begin{equation}\label{ZerothOrderField}
a_{\rm out}^0(\omega)=\frac{\Phi_0}{2\pi}\frac{1}{\bar A^*(\omega)}\sqrt{\frac{\omega}{4\pi\hbar Z_1}}\int_{-\infty}^{\infty}e^{i\omega t}\phi_0(t).
\end{equation}
We calculate now $[a_0(\omega),a_1(\omega')]+[a_1(\omega),a_0(\omega')]$,
the other cases in the leading order (between two creation operators, or between mixed operators) can be proved similarly. We get
\begin{eqnarray}
&&[a_0(\omega),a_1(\omega')]=\frac{\Phi_0}{\hbar\pi^2}\sqrt{\frac{\omega}{\omega'}}\frac{\bar A(\omega')}{\bar A^*(\omega)} \int_{-\infty}^{\infty}dtdt' \\
&& \times \int_0^{\infty}d\omega''\frac{\sin\left[\omega''(t-t')\right]}{\omega''}\cos\left[ \phi_0(t')-\omega_{\rm J}t' \right] \frac{{\rm Re}[Z_{\rm t}(\omega'')]}{R_{\rm Q}}. \nonumber
\end{eqnarray}
Here we have used that $[\phi_0(t),\sin\phi_0(t')]=-C(t-t')\cos\phi_0(t')$,
see~(\ref{eq:CommutatorCorrelator}) and the derivation below this.
We perform integration over the time $t$, use ${\rm Re}[Z_{\rm t}(\omega)]=Z_1\vert \bar A(\omega) \vert ^2$,
and obtain
\begin{eqnarray}
[a_0(\omega),a_1(\omega')]=i\frac{\Phi_0}{\hbar\pi}\sqrt{\frac{1}{\omega'\omega'}}\frac{Z_1\bar A(\omega')\bar A(\omega)}{R_{\rm Q}} \int_{-\infty}^{\infty}dt' e^{i(\omega+\omega') t'}\cos\left[ \phi_0(t')-\omega_{\rm J}t' \right]. \nonumber
\end{eqnarray}
We observe, that the result is
invariant under the change $\omega \leftrightarrow \omega'$.
It follows that $[a_0(\omega),a_1(\omega')]+[a_1(\omega),a_0(\omega')]=0$,
which is the desired property.

In the second order the calculation goes through similar steps. We calculate first
the double commutator
\begin{equation}
\left[ \phi,[\sin\phi'' , \cos\phi' ] \right] = C(t-t'')\left[ \cos\phi'' , \cos\phi'\right] +C(t-t')\left[ \sin\phi' , \sin\phi''\right].
\end{equation}
Here we use the notation $\phi=\phi_0(t)$, $\phi'=\phi_0(t')-\omega_{\rm J}t'$, and $\phi''=\phi_0(t'')-\omega_{\rm J}t''$. Using the solution
(\ref{SolutionOut2}), we get
\begin{eqnarray}
&&[a_0(\omega),a_2(\omega'')]=\frac{Z_1I_{\rm c}^2}{2\pi\hbar} \bar A(\omega)\bar A(\omega'') \sqrt{\frac{1}{\omega\omega''}} \int_{-\infty}^{\infty}dt'dt'' e^{i\omega'' t''}\times  \nonumber\\
&& \left[ 1 +\frac{{\rm Sgn}(t'-t'')}{e^{\omega_{\rm c}\vert t'-t''\vert}-1} \right] \left[ e^{i\omega t''}\left[ \cos\phi'' , \cos\phi'\right] +e^{i\omega t'}\left[ \sin\phi' , \sin\phi''\right] \right].
\end{eqnarray}
We observe, that the total part $\propto e^{i\omega t''}$
is symmetric under the exchange $\omega \leftrightarrow \omega''$. Therefore it
is canceled by the corresponding term coming from $[a_2(\omega),a_0(\omega'')]$.

The part proportional to $e^{i\omega t'}$, is equivalent to $-(1/2)\times[a_1(\omega),a_1(\omega'')]$, up to the extra part
$\propto {\rm Sgn}(t'-t'')$.
The contribution coming from $[a_2(\omega),a_0(\omega'')]$ is obtained by changing the
overall sign, and performing integration over
$e^{i\omega t''} e^{i\omega'' t'}$, instead of $e^{i\omega'' t''} e^{i\omega t'}$.
We observe symmetry with respect to $t' \leftrightarrow t''$ and $\left[ \sin\phi' , \sin\phi''\right] \leftrightarrow  \left[ \sin\phi'' , \sin\phi'\right]$:
for terms $\propto {\rm Sgn}(t'-t'')$ this expression is the opposite as the original terms
from $[a_0(\omega),a_2(\omega'')]$.
For terms not $\propto {\rm Sgn}(t'-t'')$, we have double summations of
the expression form $[a_0(\omega),a_2(\omega'')]$. Thus,
$[a_0(\omega),a_2(\omega'')]+[a_2(\omega),a_0(\omega'')]+[a_1(\omega),a_1(\omega'')]=0$,
which is the desired property.


\section*{Appendix C: Calculating averages $\left\langle e^{i\phi(t)} e^{-i\phi(t')}\right\rangle$}
\subsection*{Derivation of the term $f_{\rm t}(\omega)$}
Using the leading order solutions for both operators $a^{(\dagger)}_{\rm out}$
in the expression $\left\langle a^{\dagger}_{\rm out}(\omega) a_{\rm out}(\omega') \right\rangle$, we get a contribution for the photon flux
\begin{eqnarray}
&&\left\langle a_{1}^{\dagger}(\omega)a_{1}(\omega')\right\rangle=\frac{I_{\rm c}^2Z_1}{4\hbar\pi\sqrt{\omega\omega'}}\bar A(\omega')\bar A^*(\omega) \int_{-\infty}^{\infty}dt\int_{-\infty}^{\infty}dt'  e^{-i\omega t}e^{i\omega' t'}\times \nonumber \\
&&\left\langle e^{-i\omega_{\rm J}(t-t')} e^{i\phi_0(t)}e^{-i\phi_0(t')} +e^{i\omega_{\rm J}(t-t')} e^{-i\phi_0(t)}e^{i\phi_0(t')} \right\rangle. 
\end{eqnarray}
Here we have used the fact that expectation values
of the form $\left\langle e^{i\phi_0(t)}e^{i\phi_0(t')}\right\rangle$ are zero due
to random phase fluctuations.
Also contributions such as $\left\langle a_0(\omega)a_1(\omega')\right\rangle$
vanish due to the same reason.
We use now the following property of bosonic operators~\cite{Ingold1992},
\begin{eqnarray}
\left\langle e^{i\phi_0(t)}e^{-i\phi_0(t')} \right\rangle=\left\langle e^{-i\phi_0(t)}e^{i\phi_0(t')} \right\rangle=e^{J(t-t')},
\end{eqnarray}
where $J(t)=\left\langle [\phi_0(t)-\phi_0(0)]\phi_0(0) \right\rangle$. We then
perform a change of variables, $x=t-t'$, $y=(t+t')/2$, do integrations over $x$ and $y$, and obtain
\begin{equation}\label{OutputCorrelation}
\left\langle a_{1}^{\dagger}(\omega)a_{1}(\omega')\right\rangle=\delta(\omega-\omega') \frac{\pi I_{\rm c}^2Z_1\vert \bar A(\omega)\vert^2}{\omega}[P(\hbar\omega_{\rm J}-\hbar\omega)+P(-\hbar\omega_{\rm J}-\hbar\omega)].\nonumber
\end{equation}

\subsection*{Derivation of the term $f_{\rm in}(\omega)$}
We derive now the inelastic reflection of thermal photons, $f_{\rm in}(\omega)$. To do this we take use of the zeroth-order solution~(\ref{ZerothOrderField})
and the second-order solution~(\ref{SolutionOut2}). We get
\begin{eqnarray}
&&\left\langle a_{2}^{\dagger}(\omega)a_0(\omega'') \right\rangle =\frac{R_{\rm Q}I_{\rm c}^2}{8\pi^2}\frac{\bar A^*(\omega)}{\bar A^*(\omega'')}\sqrt{\frac{\omega''}{\omega}} \int_{-\infty}^{\infty} dt dt' dt'' \left[ 1+ \frac{{\rm Sgn}(t-t')}{e^{\omega_{\rm c }\vert t-t'\vert}-1} \right] \times \nonumber \\
&&\left\langle \left\{ \cos[\phi_0(t')-\omega_{\rm J} t']\sin[\phi_0(t)-\omega_{\rm J} t]  - {\rm H.c.}\right\} \phi_0(t'')\right\rangle \label{Derivation1}
\end{eqnarray}

The next step is to calculate the ensemble average.
By applying the Wick's theorem we get
\begin{eqnarray}
&&\left\langle e^{\pm i\phi_0(t)}e^{\mp i\phi_0(t')}\phi_0(t'')\right\rangle= \nonumber \\
&&\pm i \left[ J(t-t'')-J(t'-t'') \right] e^{J(t-t')}.\nonumber
\end{eqnarray}
We have also $\left\langle e^{\pm i\phi_0(t)}e^{\pm i\phi_0(t')}\right\rangle=0$.
These relations lead to
\begin{eqnarray}
&& \left\langle \left\{ \cos[\phi(t')-\omega_{\rm J} t']\sin[\phi(t)-\omega_{\rm J} t]  - {\rm H.c.} \right\}    \phi(t'')\right\rangle \nonumber \\
=&&\frac {1}{2}\cos[\omega_{\rm J} (t-t')]\left[ J(t-t'')-J(t'-t'') \right] \left(  e^{J(-t+t')} - e^{J(t-t')} \right)
\end{eqnarray}
We can perform integration over $t''$ by using
\begin{eqnarray}
J(t)&=&2\int_0^{\infty}\frac{d\omega}{\omega}\frac{{\rm Re}[Z_{\rm t}(\omega)]}{R_{\rm Q}} \nonumber\\
&&\times \left\{ \coth\left(\frac{1}{2}\beta\hbar\omega \right)[\cos(\omega t)-1]-i\sin(\omega t)   \right\}, \nonumber
\end{eqnarray}
and obtain
\begin{eqnarray}
&& \int_{-\infty}^{\infty} e^{i\omega'' t''}[J(t-t'')-J(t'-t'')]=2\pi \frac{ {\rm Re}[Z_{\rm t}(\omega'')]}{R_{\rm Q}\omega''}\times \nonumber\\
&&\left\{ \coth\left(\frac{1}{2}\beta\hbar\omega''\right) \left[e^{i\omega'' t}-e^{i\omega'' t'} \right]-\left[ e^{i\omega'' t}-e^{i\omega'' t'} \right] \right\} \nonumber
\end{eqnarray}
The term inside the last parentheses can be put into the form
\begin{eqnarray}
4 i\left( \frac{1}{e^{\beta\omega''-1}} \right) \sin\left(\omega''\frac{t-t'}{2}\right)\exp\left[i\omega''\frac{t+t'}{2}\right]\nonumber
\end{eqnarray}
Using these relation, we obtain
\begin{eqnarray}
&&\left\langle a_{2}^{\dagger}(\omega)a_0(\omega'') \right\rangle= i\frac{I_{\rm c}^2}{2\pi\hbar} \frac{\bar A^*(\omega)}{\bar A^*(\omega'')} \sqrt{\frac{\omega''}{\omega}} \frac{{\rm Re}[Z_{\rm t}(\omega'')]}{\omega''} \times \nonumber \\
&& \int_{-\infty}^{\infty} dt dt' e^{-i\omega t} e^{i\omega'' t''}   \cos[\omega_{\rm J} (t-t') ]\left[ e^{J(t'-t)}-e^{J(t-t')} \right] \nonumber \\
&&\times \sin\left(\omega''\frac{t-t'}{2}\right) \frac{1}{e^{\beta\hbar\omega''}-1} \left[ 1+ \frac{{\rm Sgn}(t-t')}{e^{\omega_{\rm c }\vert t-t'\vert}-1} \right] .
\end{eqnarray}

For the last two time integrations we do a change of variables $x=t-t'$ and $y=(t+t')/2$. The resulting
$y$-dependence is in a factor $\exp[iy(\omega''-\omega)]$. The
integration over $y$ leads to the factor
$2\pi\delta(\omega-\omega'')$. Performing integration over $\omega''$ and division by $2\pi$
(to obtain $f_{\rm in}$), one gets
\begin{eqnarray}
&&\int_{0}^\infty d\omega''\frac{1}{2\pi}\left \langle a_{2}^{\dagger}(\omega)a_0(\omega'') \right \rangle =\frac{2 I_{\rm c}^2}{\hbar} f_{0}(\omega)\frac{{\rm Re}[Z_{\rm t}(\omega)]}{\omega}   \times  \nonumber \\
&& \int_{-\infty}^{\infty} dx \cos(\omega_{\rm J} x) {\rm Im}[e^{J(x)}]  e^{-i\frac{x\omega}{2}}\sin\left( \frac{\omega x}{2} \right)\left[ 1+ \frac{{\rm Sgn}(x)}{e^{\omega_{\rm c }\vert x\vert}-1} \right] . \label{2ndExpression} 
\end{eqnarray}
Adding this with the contribution $\langle a_{0}^{\dagger}(\omega)a_2(\omega'') \rangle=\langle a_{2}^{\dagger}(\omega'')a_0(\omega)\rangle^*$, and using $\omega=\omega''$, one sees that only two times the real part of
(\ref{2ndExpression}) survives. Thus,
\begin{eqnarray}
&&f_{\rm in}(\omega)=f_{0}(\omega) \frac{2 I_{\rm c}^2 }{\hbar\omega } {\rm Re}[Z_{\rm t}(\omega)] \int_{-\infty}^{\infty} dx   \\
&&\times {\rm Im} [e^{J(x)}] \cos(\omega_{\rm J} x)\sin\left( \omega x \right)\left[ 1+ \frac{{\rm Sgn}(x)}{e^{\omega_{\rm c }\vert x\vert}-1} \right]. \nonumber 
\end{eqnarray}
We know that $J(-x)=J^*(x)$, and therefore ${\rm Im} [e^{J(-x)}]=-{\rm Im} [e^{J(x)}]$. Therefore the part $\propto {\rm Sgn}(x) $
cancels out due to symmetry reasons.
Because only $e^{J(x)}$ is a complex number,
the result does not change if we take the imaginary part over the whole expression [without the part $\propto {\rm Sgn}(x) $], instead of only over $e^{J(x)}$. This leads to result (\ref{eq:FluxThermalInelastic}).


\section*{Appendix D: Estimating higher-order coherences up to second order in $I_{\rm c}$}
We want to calculate expressions such
\begin{eqnarray}\label{SecondCoherence}
\left\langle a_{\rm out}^{\dagger}(\omega) a_{\rm out}^{\dagger}(\omega') a_{\rm out}(\omega'') a_{\rm out}(\omega''') \right\rangle,
\end{eqnarray}
up to second order in $I_{\rm c}$. We do this by inserting the first-order solution $a_1^{(\dagger)}$
and the zeroth-order solution $a_0^{(\dagger)}$ both twice into the four operators $a^{(\dagger)}_{\rm out}$.
We neglect the contribution if using once the second order solution $a_2^{\dagger}$, as this is proportional to $f_0(\omega)$.
We will also neglect backward directed Cooper-pair tunneling, i.e.~we take the approximation
$a_1(\omega)\propto \int dt e^{i(\omega-\omega_{\rm J}) t}e^{i\phi_0(t)}$
and neglect terms of type $\propto\int dt e^{i(\omega+\omega_{\rm J}) t}e^{-i\phi_0(t)}$.
Such tunneling against the voltage is well suppressed by the temperature.

We take use of the following result for the ensemble average of bosonic operators $\phi$,
\begin{eqnarray}\label{MixedTerms}
&&\left\langle e^{-i\phi}\phi'\phi''e^{i\phi'''}  \right\rangle=\left\langle e^{-i\phi}e^{i\phi'''}  \right\rangle\times \\
&&\left\{ \left\langle \phi'\phi'' \right\rangle +\left[  \left\langle \phi''\phi'''\right\rangle-\left\langle \phi\phi'' \right\rangle  \right] \left[  \left\langle \phi\phi' \right\rangle  -\left\langle \phi'\phi''' \right\rangle \right]   \right\}.\nonumber
\end{eqnarray}
Here we use the notation $\phi'=\phi_0(t')$ and similarly for others. The result can be derived by expressing
the exponential functions as powers series and applying the Wick's theorem.
Important here is that the order of the operators $\phi^i$ stays the same
when contracted into the pairs. This is then also
valid for permutations of the initial operators.
The result~(\ref{MixedTerms}) is also immune to exchanging the signs in the exponents.

\subsection*{Intensity cross-correlations}
We consider first the correlator
\begin{eqnarray}\label{PowerCorrelator}
\left\langle a_{\rm out}^{\dagger}(\omega) a_{\rm out}(\omega')  a_{\rm out}^{\dagger}(\omega'') a_{\rm out}(\omega''') \right\rangle.
\end{eqnarray}
Once we obtain expression for this, the other orderings of the operators $a^{(\dagger)}_{\rm out}$ can be deduced
by using general relations for the field operators, equation~(\ref{CommutationRelation}).
We need to calculate the sum of all the orderings,
\begin{eqnarray}
\left\langle e^{-i\phi}\phi'\phi''e^{i\phi'''} \right\rangle+ \left\langle e^{-i\phi}e^{i\phi'''}\phi''\phi' \right\rangle+\left\langle \phi''\phi' e^{-i\phi}e^{i\phi'''} \right\rangle + \left\langle \phi'' e^{i\phi'''}e^{-i\phi} \phi'\right\rangle\nonumber
\end{eqnarray}
However, it turns out that only the first of these four terms is important, as the other terms are proportional to $f_0(\omega)$, and can be neglected.
This can be understood by rewriting the ensemble average with the help of
a formal density matrix $\hat \rho$ of the unperturbed system,
${\rm Tr} \left\{  a(\omega')a^{\dagger}(\omega'') a(\omega''')\hat\rho a^{\dagger}(\omega)  \right\} $:
tunneling with radiation ($\propto e^{\pm i\phi_0(t)}$) has to be inserted around the density matrix, otherwise the result is zero
at $T=0$.

To calculate (\ref{PowerCorrelator}), the difficulty is to perform integration over all times of the term
\begin{equation}\label{Integration}
\left\langle e^{-i\phi}\phi'\phi''e^{i\phi'''} \right\rangle e^{i(-\omega+\omega_{\rm J})t+i\omega' t'-i\omega'' t''+i(\omega'''-\omega_{\rm J})t'''}.
\end{equation}
In result~(\ref{MixedTerms}) 
the first term inside the brackets is the easiest to calculate, as the time-dependent terms are
only functions of $t-t'''$ or $t'-t''$. We do twice similar change of variables as when calculating
$f_{\rm t}(\omega)$ (Appendix C) and obtain the first contribution for the term (\ref{Integration}),
\begin{eqnarray}
&&(2\pi)^2\int_{-\infty}^{\infty}dx_1\int_{-\infty}^{\infty}dx_2e^{J(x_1)} p(x_2) e^{ix_1(\omega_{\rm J}-\omega)}e^{ix_2\omega'}   \nonumber\\
&&\times\delta(\omega-\omega''')\delta(\omega'-\omega'')   =(2\pi)^2 \left[ 2\pi\hbar P(\hbar\omega_{\rm J}-\hbar\omega) \right]\nonumber\\
&&\times \left[4\pi\frac{1}{\omega'}\frac{{\rm Re}[Z_{\rm t}(\omega')]}{R_{\rm Q}}\right] \delta(\omega-\omega''')\delta(\omega'-\omega'').
\end{eqnarray}
Here we mark $p(t'-t'')\equiv \langle\phi(t')\phi(t'')\rangle$ and it is determined by Eq.~(\ref{EffectiveImpedance}).
We have neglected a contribution proportional to $f_0(\omega)$.

Only the last term,  proportional to $\langle\phi\phi''\rangle \langle\phi'\phi'''\rangle $, gives another finite contribution at $T=0$,
\begin{eqnarray}
&&2\pi \left[ 2\pi\hbar P(\hbar\omega_{\rm J}-\hbar\omega-\hbar\omega'') \right]  \times \left[4\pi\frac{1}{\omega'}\frac{{\rm Re}[Z_{\rm t}(\omega')]}{R_{\rm Q}}\right] \nonumber\\
&&\times \left[4\pi\frac{1}{\omega''}\frac{{\rm Re}[Z_{\rm t}(\omega'')]}{R_{\rm Q}}\right] \delta(-\omega+\omega'-\omega''+\omega''').
\end{eqnarray}
We note that in this case the operators of same type are paired [$ a(\omega')\leftrightarrow a(\omega''')$, $a^{\dagger}(\omega)\leftrightarrow a^{\dagger}(\omega'')$].
To obtain expression for (\ref{PowerCorrelator}) we sum up these two results and multiply them with $\tilde A \sqrt{\frac{\omega'\omega''}{\omega\omega'''}}\left( \frac{I_{\rm c}}{8e\pi} \right)^2$, where
\begin{equation}\label{eq:TildeA}
\tilde A=\frac{\bar A^*(\omega)\bar A(\omega''')}{\bar A^*(\omega')\bar A(\omega'')}.
\end{equation}
We obtain for the intensity cross correlations up to second order in $I_{\rm c}$,
\begin{eqnarray}\label{SecondCoherence2}
&& 2\pi\frac{I_{\rm c}^2}{2\omega}P(\hbar\omega_{\rm J}-\hbar\omega){\rm Re}[Z_{\rm t}(\omega)]\delta(\omega-\omega''')\delta(\omega'-\omega'')+\nonumber\\
&& \tilde A\delta(\omega-\omega'+\omega''-\omega''') \frac{2\pi I_{\rm c}^2}{\sqrt{\omega\omega'\omega''\omega'''}} \times\\
&& P[\hbar(\omega_{\rm J}-\omega-\omega'')]\left\{\frac{ {\rm Re}[Z_t(\omega')]{\rm Re}[Z_t(\omega'')]}{R_{\rm Q}}\right\}.\nonumber
\end{eqnarray}

\subsection*{Second-order coherence $G^{(2)}$}
We consider now the  expectation value
$
\left\langle  a^{\dagger}(\omega)a^{\dagger}(\omega')a(\omega'')a(\omega''') \right\rangle
$.
By using the identity
$[a_{\rm out}(\omega),a_{\rm out}^{\dagger}(\omega')]=\delta(\omega-\omega')$ and doing the exchange
$\omega'\leftrightarrow\omega''$ in term~(\ref{SecondCoherence2}),
we get the result
\begin{eqnarray}
&&\left\langle  a^{\dagger}(\omega)a^{\dagger}(\omega')a(\omega'')a(\omega''') \right\rangle=\nonumber \\
&&  \tilde A(\omega,\omega'',\omega',\omega''')\delta(\omega+\omega'-\omega''-\omega''') \frac{2\pi I_{\rm c}^2}{\sqrt{\omega\omega'\omega''\omega'''}} \times \nonumber \\
&& P[\hbar(\omega_{\rm J}-\omega-\omega')]\left\{\frac{ {\rm Re}[Z_{\rm t}(\omega')]{\rm Re}[Z_{\rm t}(\omega'')]}{R_{\rm Q}}\right\}. \label{BunchingGeneral} 
\end{eqnarray}
This can also be derived through a direct calculation, as was done for the
intensity cross correlator.

\subsection*{Cauchy-Schwarz inequality}
The Cauchy-Schwarz inequality compares intensity correlations
with the second-order coherence. We calculate these in a small frequency interval $\Delta\omega$
around the frequencies $\omega_a$ and $\omega_b$. This assumes a filtering of the measured output signal
into these frequencies (intensity-correlations).

We integrate result (\ref{BunchingGeneral}) over four frequencies, each of them having the interval $\omega\in[\omega_a-\Delta\omega/2,\omega_a+\Delta\omega/2]$. We get by assuming a small $\Delta\omega$,
\begin{equation}
G^{(2)}(0)\approx  \frac{2\pi I_{\rm c}^2 \Delta\omega^3}{\omega_a^2 R_{\rm Q}}P\left[ \hbar(\omega_{\rm J}-2\omega_a) \right]\left[{\rm Re}\left[ Z_{\rm t}(\omega_a) \right]\right]^2.
\end{equation}
Here we have used $\tilde A(\omega_a,\omega_a,\omega_a,\omega_a)=1$.
Similarly for the contribution at $\omega_b$.

The cross correlations between the two frequencies, $\omega_a-\omega_b\gg \Delta\omega$, give
\begin{eqnarray}
&&\int_{\omega_a-\Delta\omega/2}^{\omega_a+\Delta\omega/2}d\omega d\omega' \int_{\omega_b-\Delta\omega/2}^{\omega_b+\Delta\omega/2}d\omega''  d\omega''' \left\langle a^{\dagger}(\omega)a (\omega')a^{\dagger}(\omega'')a(\omega''')\right\rangle\\
&& \approx  \tilde A(\omega_a,\omega_a,\omega_b,\omega_b)
\frac{2\pi I_{\rm c}^2 \Delta\omega^3}{\omega_a\omega_b R_{\rm Q}} P\left[ \hbar(\omega_{\rm J}-\omega_a-\omega_b) \right] {\rm Re}\left[ Z_{\rm t}(\omega_a) \right] {\rm Re}\left[ Z_{\rm t}(\omega_b) \right] .\nonumber
\end{eqnarray}
We notice from (\ref{eq:TildeA}) that also $\tilde A(\omega_a,\omega_a,\omega_b,\omega_b)=1$.
(Actually an additional factor 2/3 appears for both cross- and autocorrelations,
due to the specific form of the bandwidth cutoff, but is neglected here for simplicity.)

\subsection*{Bunching for $Z_0=Z_1$}
For the free-space configuration ($Z_0=Z_1$) we have
a simple result $\tilde A={\cal C}^*(\omega'){\cal C}(\omega'')/{\cal C}^*(\omega){\cal C}(\omega''')$.  In the following we will assume
that $\omega_{\rm J}\ll\omega_{\rm c}$, so that $\tilde A=1$ and ${\rm Re}[Z_{\rm} (\omega)]=Z_0$.
One obtains for the second-order coherence (when integrated over all frequencies)
\begin{eqnarray}
&& G^{(2)}(0)=\frac{2\pi}{R_{\rm Q}} \left( \frac{I_{\rm c} \hbar Z_0^2}{4\pi} \right)^2 \int_{0}^{\infty}d\omega\int_{0}^{\infty}d\omega'\int_{0}^{\omega+\omega'} d\omega'' \nonumber \\
&&\times P[\hbar(\omega_{\rm J}-\omega-\omega')]= \frac{2\pi}{R_{\rm Q}} \left( \frac{I_{\rm c} \hbar Z_0^2}{4\pi} \right)^2 \label{G2General} \\
&&\times\int_0^{\infty}d\omega\int_0^{\infty}d\omega'(\omega+\omega')P[\hbar(\omega_{\rm J}-\omega-\omega')].\nonumber
\end{eqnarray}
We do now a change of variables: $s=\omega+\omega'$, $t=(\omega-\omega')/2$, and get
\begin{eqnarray}
&& G^{(2)}(0)= \frac{2\pi}{R_{\rm Q}} \left( \frac{I_{\rm c} \hbar Z_0^2}{4\pi} \right)^2\int_0^{\infty}d s\int_{-s/2}^{s/2}dt sP[\hbar(\omega_{\rm J}-s)] \nonumber \\
&&=\frac{2\pi}{R_{\rm Q}} \left( \frac{I_{\rm c} \hbar Z_0^2}{4\pi} \right)^2\int_0^{\infty}d s s^2P[\hbar(\omega_{\rm J}-s)].\label{eq:SecondOrderOpen}
\end{eqnarray}
On the other hand, in the same approximation the first-order coherence is
\begin{equation}
G^{(1)}_{\rm t}(0)=\frac{I_{\rm c}^2Z_0^2}{4}\frac{e^{-2\gamma\rho}}{\Gamma(1+2\rho)}\left(\frac{\pi\rho\hbar\omega_{\rm J}}{4E_C}\right)^{2\rho}.
\end{equation}
This gives
\begin{eqnarray}\label{eq:bunching}
g^{(2)}_{\rm t}(0)=\left( \frac{\hbar\omega_{\rm J} }{\pi E_{\rm J}} \right)^2\left[ \pi\rho\frac{\hbar\omega_{\rm J}}{4E_C} \right]^{-2\rho} \frac{\Gamma(1+2\rho)e^{2\gamma\rho}}{1+3\rho+2\rho^2}.
\end{eqnarray}
For small $\rho$ we get $g^{(2)}_{\rm t}(0)=(\hbar\omega_{\rm J}/\pi E_{\rm J})^2$.
For a general time $\tau$ we substitute
\begin{equation}
s^2\rightarrow \left( \frac{2}{\tau} \right)^2\sin^2\left( \frac{\tau s}{2} \right),
\end{equation}
in equation (\ref{eq:SecondOrderOpen}).

Let us consider restricted region of frequency interval, $\omega_0-\Delta\omega/2<\omega<\omega_0+\Delta\omega/2$, for all frequencies. We get a new integration range,
\begin{eqnarray}
&& \int_{2\omega_0}^{2\omega_0+\Delta\omega} ds \int_{s/2-\omega_0-\Delta\omega/2}^{-(s/2-\omega_0)+\Delta\omega/2} dt\int_{s-\omega_0-\Delta\omega/2}^{\omega_0+\Delta\omega/2} d\omega'' +\nonumber\\
&& + \int_{2\omega_0-\Delta\omega}^{2\omega_0} ds \int^{s/2-\omega_0+\Delta\omega/2}_{-(s/2-\omega_0)-\Delta\omega/2} dt\int_{\omega_0-\Delta\omega/2}^{s-\omega_0+\Delta\omega/2} d\omega'' . \nonumber
\end{eqnarray}
Since the integrant (\ref{G2General}) is independent of both $\omega''$ and $t$, the integration range gets the form
\begin{eqnarray}
\int_{2\omega_0-\Delta\omega}^{2\omega_0+\Delta\omega} ds \left( -\vert 2\omega_0-s \vert+\Delta\omega \right)^2, \nonumber
\end{eqnarray}
and one obtains for the corresponding second-order coherence
\begin{eqnarray}
&& G_{\rm t}^{(2)}(0) = \frac{2\pi}{R_{\rm Q}} \left( \frac{I_{\rm c}\hbar Z_0^2}{4\pi} \right)^2 \\ &&\int_{2\omega_0-\Delta\omega}^{2\omega_0+\Delta\omega} ds \left( -\vert 2\omega_0-s \vert+\Delta\omega \right)^2  P[\hbar(\omega_{\rm J}-s)]. \nonumber
\end{eqnarray}
For low temperatures we have $P(E)\approx 0$ for $E<-k_{\rm B}T$.
Therefore the result is practically zero if the lower limit of integration is above $\omega_{\rm J}$. Optimal result is obtained for
integration range that covers symmetrically the $P(E)$ peak at $E=0$, i.e.~for $\omega_0=\omega_{\rm J}/2$.
For small $\Delta\omega$, but still larger than the thermal width $\Gamma$, the $P(E)$-function
can be approximated as $\delta(E)$, and the integration gives in this case ($\rho\ll 1$, $\Delta\omega\gg \Gamma$)
\begin{eqnarray}
G_{\rm t}^{(2)}(0) = \frac{2\pi\hbar}{ R_{\rm Q}} \left( \frac{I_{\rm c} Z_0^2}{4\pi} \right)^2 \Delta\omega^2=  \left(\frac{I_{\rm c}^2Z_0^2}{4}\right)^2  \left(\frac{\hbar \Delta\omega}{\pi E_{\rm J}}\right)^2 . \nonumber
\end{eqnarray}
This leads to result (\ref{eq:BunchingSmallInterval}).


\bibliographystyle{plainnat}

\end{document}